\documentclass{elsarticle}

\usepackage{lineno,hyperref}
\usepackage[cmex10]{amsmath}
\usepackage{amsfonts}
\usepackage{floatrow}
\usepackage[singlelinecheck=off]{subfig}
\usepackage{caption}
\usepackage{tabu}
\usepackage{longtable}
\newcommand{\ud}{\,\mathrm{d}}

\journal{Signal Processing}

\makeatletter
 \renewenvironment{table}
     {\@float{table} \small}
     {\end@float}
 \makeatother

\begin{document}

\begin{frontmatter}

\title{Characterisation and extraction of a Rayleigh-wave mode in vertically-heterogeneous media using Quaternion SVD}
\tnotetext[mytitlenote]{Part of this work was presented at the 73rd EAGE Conference and Exhibition incorporating SPE EUROPEC 2011.}

%% or include affiliations in footnotes:
\author[mymainaddress]{Angelo Sajeva\corref{mycorrespondingauthor}}
\ead{angelo.sajeva@for.unipi.it}

\author[mysecondaryaddress]{Giovanni Menanno}
\cortext[mycorrespondingauthor]{Corresponding author}
%\ead{support@elsevier.com}

\address[mymainaddress]{Earth Sciences Department, University of Pisa, Via Santa Maria, 53, 56126, Pisa}
\address[mysecondaryaddress]{Formerly at Earth Sciences Department, University of Pisa}

\begin{abstract}
We propose a method that identifies a mode of Rayleigh waves and separates it from body waves and from other modes, using quaternions to represent multi-component data. Being well known the abilities of quaternions to handle rotations in space, we use previous results derived from Le Bihan and Mars \cite{LeBihan04} to prove that a Rayleigh-wave mode recorded by an array of vector-sensors can be approximated by a sum of trace-by-trace rotating time signals. Our method decomposes the signal into narrow-frequency bands, which undergo both a velocity correction and a polarisation correction. The aim of these corrections is to reduce the mode of interest to a quasi-monochromatic wave packet with infinite apparent velocity and quasi-circular polarisation. Once written in quaternion notation, we refer to this wave packet as ``quaternion brick''. Based on theoretical considerations, we prove that this quaternion brick maps into the first quaternion eigenimage of the Quaternion Singular Value Decomposition. We apply this method to synthetic datasets derived from two vertically-heterogeneous models to extract the fundamental mode and we prove that it is correctly separated from either a higher mode of propagation or body waves with negligible residual. Results are presented in both time-offset and frequency-phase slowness domains.
\end{abstract}


\end{frontmatter}

%\linenumbers

\section{Introduction}

Multi-component signal processing is experiencing growing interest in various research areas such as seismic prospecting \cite{Anderson1996}, wireless communication \cite{Andrews2001}, and electromagnetism \cite{Nehorai1994a}. The acquisition of multi-component signals is carried out by means of vector-sensors, which record the three components of vibrations propagating through a three-dimensional medium. In the case of elastic waves the vector-sensor is composed of co-located directional geophones which record the particle displacements along three orthogonal directions.

In seismology and seismic prospecting, one of the benefits of recording the seismic wavefield with vector-sensors is the possibility of classifying seismic waves based on their polarisation. For instance, in vertically-heterogeneous media, Rayleigh waves (a type of surface waves characterised by elliptical polarisation in the vertical plane defined by the direction of motion of the wave), may be distinguished from Love waves (another type of surface waves whose polarisation is linear in the transverse-horizontal direction) and pressure and shear waves (body waves characterised by linear polarisation). Moreover, in the literature, several publications show applications of data-filtering and wavefield-separation techniques based on the polarisation information \cite{diallo2006instantaneous,Donno2008,ShiehHermann1990,tiapkina2012single}.

The practical need of separating Rayleigh waves from body waves in seismic acquisitions is one of the objectives of the present work. The separation between Rayleigh waves and body waves is of general interest in both deep-exploration and near-surface geophysics{. In exploration geophysics, Rayleigh waves, usually referred to as} ground roll, are seen as noise to be rejected from data. Differently, in near-surface geophysics, the identification and extraction of Rayleigh-wave modes can help the inversion procedure for the estimation of the shear-wave velocity profile, which is a key parameter for site characterisation \cite{socco2004surface}. 

In near-surface geophysics, early methods, such as the SASW (Surface Analysis of Surface Waves), limited the analysis of surface waves to the sole fundamental mode \cite{Stokoe1983}. Nevertheless, higher normal-modes of the Rayleigh waves are often present in the data and may mislead the interpretation of the dispersive characteristics. Even in the case of multiple receivers, as in the MASW (Multi-channel Analysis of Surface Waves) experiments \cite{Park1999}, the identification of different modes may be difficult due to osculation points between modes \cite{Forbriger2003}. For these reasons, another objective of this paper is the separation between interfering Rayleigh-wave modes. Karray and Lefebvre \cite{Karray2009} proposed to separate the contribution of two interfering modes of the Rayleigh wave on single-component data by suppressing one of them using a time-variable filter, being known the dispersion characteristics of the other one. However, the availability of multi-component seismic data, such as those acquired using triaxial sensors, creates novel possibilities for the detection and separation of surface waves that consider the whole vectorial wavefield.

%[11] \cite{Nazarian1984}
%[15,16] \cite{Knopoff1964,Landisman1969}

A classical approach in vector-sensor signal processing is based on concatenating the spatial components into a long vector before processing \cite{Anderson1996,Nehorai1994a}. In this paper, we use a different approach which consists in using quaternion numbers to represent the vectorial time series recorded by vector sensors.
Quaternions are numbers with a real part and three imaginary components (hypercomplex numbers). Their behavior is described by quaternion algebra, an extension of complex algebra introduced by Hamilton \cite{hamilton1844}, which is non-commutative. Quaternion algebra has been used in various scientific fields, ranging from theoretical physics \cite{ALD_1995} to robotics and computer graphics \cite{PLTN_1989}. In seismic processing, it has been used to perform various tasks, such as wavefield separation \cite{LeBihan04}, polarisation studies \cite{MIR_LBH_MAR_2006}, multi-component velocity analysis \cite{GRA_MAZ_STU_2007}, multi-component deconvolution \cite{MEN_MAZ_2012}, vector interpolation of multi-component data \cite{StantonSacchi2013}, and re-orientation of vector-sensors \cite{Krieger2015}.
To our knowledge, Le Bihan and Mars \cite{LeBihan04} were the first to introduce quaternion algebra to seismic data processing, proposing a method to separate wavefields in multi-component multi-channel seismic acquisitions based on an extension to the quaternion field of Singular Value Decomposition (QSVD) \cite{Zhang1997}. 
In this work, we propose a variation of the method of Le Bihan and Mars \cite{LeBihan04} that permits to separate a broad-band Rayleigh-wave mode from body waves and from higher-order Rayleigh-wave modes using the properties of QSVD. Results of a preliminary work on this subject have been presented in \cite{Sajeva2011}, and further developments in \cite{Sajeva2016}. 

The paper is organised as follows. First, we briefly introduce quaternions and their properties. Next, we describe the manipulations applied to the data that enable a compact description of a Rayleigh-wave mode using quaternions and we present the extraction method. Finally, we apply this method to synthetic data and we demonstrate that it is able to distinguish and separate a Rayleigh-wave mode from both body waves and higher-order Rayleigh-wave modes.

\section{Brief introduction to Quaternions} \label{se:quat}

Quaternions are hypercomplex numbers introduced by Hamilton \cite{hamilton1844} in the attempt of generalising the field of complex numbers to the three-dimensional space. A quaternion $q$ is defined by three imaginary units $\mathtt{i},\mathtt{j}$, and $\mathtt{k}$ and can be written as:
% EQ 1
\begin{align}
q &= a+\mathtt{i}b+\mathtt{j}c+\mathtt{k}d, \qquad a,b,c,d \in \mathbb{R} \label{eq:quat}.
\end{align}
The multiplication rules of the imaginary units are:
% EQ 2
\begin{align} 
\mathtt{ii} &= \mathtt{jj} = \mathtt{kk}= -1 \nonumber \\ 
\mathtt{ij} &= -\mathtt{ji}= \mathtt{k} \quad \mathtt{jk}=-\mathtt{kj} = \mathtt{i} \quad \mathtt{ki} = - \mathtt{ik} = \mathtt{j}. \label{eq:vectprod}
\end{align} 
Consequently, the multiplication in quaternion algebra is noncommutative. Note that the set of quaternions $\mathbb{H}$ is isomorph to $\mathbb{R}^4$, thus, it can be treated as a vector space over the real numbers. In $\mathbb{H}$, a number of operations such as the inverse operation with respect to multiplication, the complex conjugation, the norm and others are defined by analogy with the complex field \cite{Zhang1997}. 
	 	 	
A point in the Euclidean space can be represented by a pure quaternion, i.e., a quaternion with null real part:
% EQ 3 (da mettere)
\begin{align} 
\mathbf{u} = (u_x,u_y,u_z)^T \iff \mathring{u} = 0 + { \tt i}u_x + { \tt j}u_y + { \tt k}u_z. \label{eq:vector_quat}
\end{align}
A rotation in space of a point $\mathring{u}$ by an angle $\phi$ around an axis $\omega$, can be written in $\mathbb{H}$ as:
\begin{equation} \label{eq:quaternion_rotation_rules_original}
\mathring{u}' = \mathring{r}_{\phi/2}\mathring{u}\mathring{r}^*_{\phi/2},
\end{equation} 
where $\mathring{u}$ is the quaternion equivalent of $\mathbf{u}$, $^*$ is the complex-conjugate operator (which inverts the sign of the imaginary parts), and $\mathring{r}_{\phi/2}$ is a unit quaternion (with norm equal to one) that embeds the information about the angle and the axis of the rotation. In the special case in which the point $\mathring{u}$ is constrained to move in a plane, and the rotation axis is perpendicular to this plane, it holds that:
% EQ 4
\begin{equation} \label{eq:quaternion_rotation_rules}
\mathring{u} \mathring{r}^*_{\phi} = \mathring{r}_{\phi} \mathring{u}. 
\end{equation}
See \ref{app:quaternion_rotation} for more details on the representations of rotations with quaternions and for a proof of Eq.~\eqref{eq:quaternion_rotation_rules}.

\subsection{Rank and QSVD of quaternion matrices}

For matrices with quaternion entries, the rank can be defined by analogy with the real and complex cases as the maximum number of columns that are right linearly independent \cite{Zhang1997}. We write the Singular Value Decomposition of a matrix $\mathbf{\mathring{A}} \in \mathbb{H}^{N \times M}$ with quaternion entries in the form:
% EQ 5
\begin{align} \label{eq:define_qsvd}
\mathbf{\mathring{A}}=\sum_{i=1}^r \boldsymbol{\mathring{\mathcal{A}}}_i, 
\end{align}
where $r$ is the rank of the matrix ($r <= min(N,M)$) and each $\boldsymbol{\mathring{\mathcal{A}}}_i \in \mathbb{H}^{N \times M}$ is a quaternion rank-1 matrix also called \emph{eigenimage}, which is given by:
% EQ 6
\begin{align} \label{eq:qsvd}
\boldsymbol{\mathring{\mathcal{A}}}_i= \sigma_i \mathbf{\mathring{w}}_i \mathbf{\mathring{v}}_i^\dagger,
\end{align}
where $\sigma_i$ is the $i$-th real singular value of the $\mathbf{\mathring{A}}$ matrix, $\mathbf{\mathring{w}}_i$° and $\mathbf{\mathring{v}}_i$ are the $i$-th columns of two quaternion unit matrices, and $^\dagger$ denotes the transposed quaternion conjugate. An important theorem states that the rank of a quaternion matrix equals the number of non-zero singular values in the matrix \cite{Zhang1997}. Consequently, a rank-1 quaternion matrix has a single non-zero singular value and, vice versa, a quaternion matrix having a single non-zero singular value is a rank-1 quaternion matrix. 

Accordingly, we can imagine the $\boldsymbol{\mathring{\mathcal{A}}}_i$ as a base of the space of quaternion matrices, i.e., each $\boldsymbol{\mathring{\mathcal{A}}}_i$ represents a direction in this space. Note that each $\boldsymbol{\mathring{\mathcal{A}}}_i$ embodies a singular value $\sigma_i$ which weights the contribution of the $i$-th direction to the $\mathbf{\mathring{A}}$ matrix. Because by definition the singular values are sorted in descending order, the $\boldsymbol{\mathring{\mathcal{A}}}_i$ terms in the summation become smaller and eventually negligible as $i$ increases. Thus, it is possible to approximate the $\mathbf{\mathring{A}}$ matrix by truncating the sum after a given $i'$ value and by dropping every eigenimage $\boldsymbol{\mathring{\mathcal{A}}}_i$ such that $i > i'$. This approximation is at the base of the subspace method extended to quaternion algebra by \cite{LeBihan04}.

\subsection{Representation of seismic data using quaternions}

Quaternions are well suited to represent multi-component seismic data \cite{Mazzotti2012}. The three components of the time series $u_x[t]$, $u_y[t]$, and $u_z[t]$ can be written as a single vector of pure quaternions $\mathring{u}[t]$ and an array of traces (a seismic gather) can be written as a matrix with pure quaternion entries:
% EQ 7
\begin{equation} \label{eq:quat_mat}
\mathring{u}[x,t] := \mathbf{\mathring{U}} = { \tt i} \mathbf{U}_x + { \tt j} \mathbf{U}_y+ { \tt k} \mathbf{U}_z,
\end{equation}
where $\mathbf{U}_x, \mathbf{U}_y,$ and $\mathbf{U}_z \in \mathbb{R} ^{N \times M}$ are three matrices with real entries that are functions of the offset, $x$, and the time, $t$, and describe the ground motion on the surface in the $x$, $y$, and $z$ directions, respectively.

\section{Quaternion extraction of a Rayleigh-wave mode} \label{se:Q_extraction}

The objective of this section is to demonstrate that the particle displacements of a quasi-monochromatic Rayleigh-wave mode, propagating on the surface and recorded by an array of vector sensors, can be written as a rank-1 quaternion matrix. 
Similar concepts have been introduced by Le Bihan and Mars \cite{LeBihan04}. In their paper, the authors note that for dispersive signals recorded on equally-spaced sensors, after a velocity correction, the difference between the phase and group velocity corresponds to a constant phase shift from one trace to another. They show that quaternions are especially capable of compactly representing such phase-shifted signals using low-rank quaternion matrices. 
We start from their analysis and we extend it to the case of a broadband signal, in which the frequency-dependent velocity and polarisation are taken into account. In particular, we discuss the effects of group velocity and polarisation corrections on a narrow-band Rayleigh-wave mode. These modifications allow us to compactly represent the dispersive waves with quaternions and to derive our extraction procedure, which relies on accurate velocity and polarisation corrections on narrow frequency bands.

For the purposes of this paper, the group velocity dispersion curve and the polarisation characteristics are assumed to be known for each frequency. In other words, we use the theoretical values instead of values estimated from the data to modify the narrow-band wave. This is done because we are mainly interested in studying the ability of quaternions to compactly represent the signal of interest and we want to minimise the error due to poor group velocity and polarisation corrections. For the interested reader, methods to estimate these parameters are discussed in Section \ref{se:appl-xmpl-single}. 

\subsection{Quasi-monochromatic Rayleigh-wave mode}

Integral expressions using a matrix formulation have been developed to describe the $x$ and $z$ particle displacements induced by a monochromatic Rayleigh wave mode \cite{Harkrider1964,Tokimatsu1992}. As we aim to describe the dispersion characteristics of the Rayleigh-wave mode in terms of the group velocity, which is a concept attached to a wave packet having a continuous spectrum, we are interested in a solution for a narrow frequency band $[\omega_{0} - \Delta \omega, \omega_0 + \Delta \omega]$. Following \cite{AkiRichards02}, it is possible to derive an approximated formula for the $x$ and $z$ particle displacements integrating the single-mode solution with unit amplitude and zero initial phase over a finite frequency band around a given angular frequency $\omega_0$. Therefore, the $x$ and $z$ particle displacements are:
% EQ 8
\begin{align} \label{eq:stationary_phase_equations}
u_x(x,t) &= r_x \Delta \omega \frac{\sin Y}{Y} 
 \cos \left[ \omega_0 \left( t - \frac{x}{v_{ph}} \right) \right] \nonumber \\
u_z(x,t) &= r_z \Delta \omega \frac{\sin Y}{Y} 
 \sin \left[ \omega_0 \left( t - \frac{x}{v_{ph}} \right) \right], 
\end{align}
where $r_x$ and $r_z$ are real-valued functions of frequency $\omega$, $v_{ph}$ is the phase velocity at frequency $\omega_0$, and $Y$ is:
% EQ 9
\begin{align} \label{eq:Y_function}
Y &= \frac{\Delta \omega}{2} [t - ( \ud k / \ud \omega)_{\omega_0} x]. 
\end{align}
Note that both components oscillate with the central frequency, are modulated by an envelope with the shape of a cardinal sine, and propagate with group velocity $v_g = 1/( \ud k / \ud \omega)_{\omega_0}$. In addition, there is a 90-degree phase shift between $u_x$ and $u_z$. This indicates that the particle orbit is elliptical in the vertical plane containing the direction of propagation of the wave packet, and that the axes of the ellipse of polarisation coincide with the vertical and horizontal directions. Also note that in Eq. \eqref{eq:stationary_phase_equations} we use the approximations $v_{ph}(\omega) \approx v_{ph}$ and $v_{g}(\omega) \approx v_{g}$, that may be inaccurate for large $\Delta \omega$ or steep dispersion curves.

In a vertically heterogeneous medium, multiple modes of Rayleigh waves propagate. Only one mode (the fundamental one) propagates for all frequencies whereas the $i$-th higher modes propagate for frequencies higher than a cut-off frequency $\omega_i$ which increases with $i$ \cite{AkiRichards02}.
How many and which modes are excited in a seismic acquisition depends on both the soil stratification and the frequency and depth of the seismic source \cite{Gucunski1991,Muller}.
In this paper, we will concentrate on the extraction of a single mode of the Rayleigh wave and we will consider all other modes as noise to be removed.

\subsection{Infinite Group-Velocity Correction}\label{se:inf-grp-vel-corr}

To exploit the ability of the SVD to separate an aligned signal from other misaligned signals or incoherent noise \cite{KIR_DONE_1999}, we apply a velocity correction that aligns the quasi-monochromatic Rayleigh-wave mode of Eq. (\ref{eq:stationary_phase_equations}). Knowing $v_g$ at frequency $\omega_0$, we can apply a time-axis transformation $t'=t- {x}/{v_{g}}$.
After this transformation, the wave packet is horizontally aligned in the time-offset domain. We will refer to this operation as infinite-group velocity correction throughout this paper. After this correction, the displacement components of Eq. (\ref{eq:stationary_phase_equations}) become
% EQ 12
\begin{align} \label{eq:displacement_single_frequency}
\hat{u}_x(x,t') &= R_x(t')
\cos ( \omega_0 t' - \phi(x) ) \nonumber \\
\hat{u}_z(x,t') &= R_z(t')
\sin ( \omega_0 t' - \phi(x) ), 
\end{align}
where:
\begin{align} \label{eq:variables_for_displ_sing_freq}
\phi(x) &= \omega_0 x \left( \frac{1}{v_{ph}} - \frac{1}{v_{g}} \right) %\nonumber \\
%XXX GM1 R_x(t') &= r_x \Delta \omega \frac{\sin Y}{Y} XXX GM1\nonumber \\
%XXX GM1 R_z(t') &= r_z \Delta \omega \frac{\sin Y}{Y} XXX GM1\nonumber \\
\end{align}%} 
and $R_x$ and $R_z$ have been introduced to compactly denote the evanescent amplitude terms that modulates the amplitude of the sinusoids oscillating at $\omega_0$ angular frequency (harmonic terms). Note that in Eq. (\ref{eq:displacement_single_frequency}), the spatial dependence affects the harmonic terms alone, and it corresponds to a phase shift $\phi(x)$.

Although this formulation is correct for any kind of disposition of aligned receivers at surface, e.g. irregularly sampled, multiple receivers or just a couple of them, like in the SASW method, we focus on the case of an array of equally-spaced receivers at positions
% EQ 14
\begin{equation} \label{eq:xcondition}
x_m = (m-1) \Delta x + x_0, \qquad m=1, 2, ..., M.
\end{equation}
Being the time domain also regularly sampled, the data can be represented as a $N \times M$ matrix, subjected to the sampling theorem for what concerns time and spatial aliasing and resolution. In this situation, each pair of adjacent traces at positions $x_m$ and $x_{m+1}$, are equal apart from the phase shift
% EQ 15
\begin{equation} \label{eq:constant_phase_shift}
\Delta \phi \approx \omega_0 \Delta x \left( \frac{1}{v_{ph}} - \frac{1}{v_g} \right).
\end{equation}
Note that the phase shift is caused by the difference between the wave-packet velocity (group velocity) and the peaks' velocity (phase velocity), i.e. it is a consequence of frequency dispersion that always characterise Rayleigh waves in real media (only Rayleigh waves propagating over an infinite half-space are not dispersive). For non-dispersive signals, the two velocities match and the phase shift is null. 

\subsection{Correction of the instantaneous polarisation} \label{se:polarization}

According to \cite{LeBihan04}, rotations in the 3D space can represent a phase shift of a signal. In this section we show that in order to represent a single narrow-band Rayleigh-wave mode as a trace-by-trace rotating signal it is necessary to perform an amplitude correction that transforms the instantaneous elliptical polarisation into a circular one. 

We have already observed that the quasi-monochromatic dispersive signal shown in Eq. (\ref{eq:displacement_single_frequency}) exhibits an instantaneous elliptic polarisation for a given receiver position and time. In fact, it holds that:
% EQ 16
\begin{equation} \label{eq:instantaneous_elliptic_polarization}
 \frac{\hat{u}_x^2}{R_x^2} + \frac{\hat{u}_z^2}{R_z^2} = 1,
\end{equation}
where $R_x$ and $R_z$ are the amplitudes of the axes of the instantaneous ellipse of polarisation. Because $R_x$ and $R_z$ vary in time, the overall polarisation of the wave packet appears to be spiral-shaped rather than elliptic.
To satisfy the condition of circular polarisation in the general case, we should perform a very accurate correction for $R_x(t')$ and $R_z(t')$ time-sample per time-sample. However, in this paper we rely on two assumptions for $R_x(t')$ and $R_z(t')$ that greatly simplify the correction procedure:

\begin{enumerate}[1.]
\item $R_x(t')$ and $R_z(t')$ vary slowly compared to the ellipse time period,
\item the lengths of the axes satisfy the relation $R_z \approx c R_x, $
where $c$ is the real-valued vertical-to-horizontal ratio of the $\omega_0$ monochromatic signal. 
\end{enumerate}

The first assumption is reasonable if $\Delta \omega$ is sufficiently small and implies that the polarisation of the mode can be approximated to be elliptic in a single time period $2\pi/\omega_0$. The second assumption is known to be exact for monochromatic signals (the vertical-to-horizontal spectral ratio can be theoretically computed) and consequently, it is a good approximation for sufficiently small $\Delta \omega$. In the hypothesis that these approximations are correct, it suffices to multiply the $x$ displacement component by $c$ to obtain an approximately circular polarisation:
% EQ 18
\begin{align} \label{eq:circularized_displacement}
\underline{u}_x &= c \hat{u}_x \nonumber \\
\underline{u}_z &= \hat{u}_z 
\end{align}
Throughout the paper, this operation is referred to as circularisation. Note that at the frequencies for which the Rayleigh-wave mode exhibits linear polarisation (for instance, when it moves from prograde to retrograde motion) the polarisation cannot be reduced to a circular one.

\subsection{Phase-shift as trace-by-trace rotation} \label{se:polarization_rot}

We introduce a new notation for simplicity, that is:
% EQ 18b NUOVA angelo TRACES NOTATION
\begin{align} \label{eq:u_xm}
\mathbf{\underline{u}}_{x,m} &:= \left\{ \underline{u}_{x}(x=x_m,t'=(n-1) \Delta t) \right\}_{n=1,..., N} \qquad m=1, 2, ..., M \nonumber \\
\mathbf{\underline{u}}_{z,m} &:= \left\{ \underline{u}_{z}(x=x_m,t'=(n-1) \Delta t)\right\}_{n=1,..., N} \qquad m=1, 2, ..., M,
\end{align}
in which also the time domain is discretised. We observe that, in the case of circular polarisation and equidistant geophones, the signal recorded by the $(m+1)$-th geophone $(\mathbf{\underline{u}}_{x,m+1}, \mathbf{\underline{u}}_{z,m+1})^T$ is the rotated version of the signal recorded by the $m$-th geophone $(\mathbf{\underline{u}}_{x,m}, \mathbf{\underline{u}}_{z,m})^T$ and in matrix notation is
% EQ 19
\begin{equation} \label{eq:rotation_between_columns}
\begin{pmatrix} \mathbf{\underline{u}}_{z,m+1}^{T} \\ \mathbf{\underline{u}}_{x,m+1}^{T} \end{pmatrix} =
P(\Delta \phi)
\begin{pmatrix} \mathbf{\underline{u}}_{z,m}^{T} \\ \mathbf{\underline{u}}_{x,m}^{T} \end{pmatrix} , \qquad m=1,...,M-1,
\end{equation}
where $P(\Delta \phi)$ is the 2D rotation matrix. The proof that the signal in Eq. (\ref{eq:circularized_displacement}) verifies Eq. (\ref{eq:rotation_between_columns}) is given in \ref{app:phase_shift_as_rotation}.

\begin{figure} %figure1 (pdf)
  \begin{center}
    \includegraphics[scale=1]{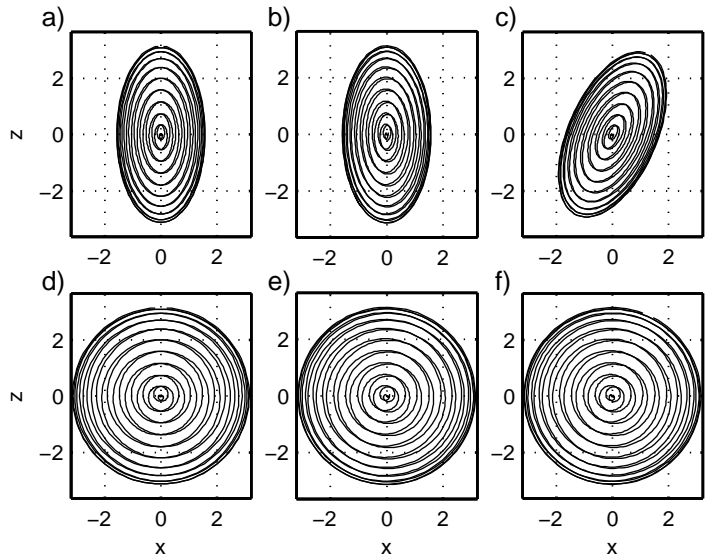} 
    \caption{(a) Hodogram of the signal of Eq. (\ref{eq:stationary_phase_equations}) using $x$=0,$\Delta \omega$=$2\pi*0.5$ rad/s, $\omega_0$=$2\pi*5$ rad/s, $r_x$=$0.5$, and $r_z$=$1$. Note that this signal has instantaneous elliptical polarisation but an overall spiral-shaped polarisation; (b) the elliptically polarised signal after a phase-shift given by the difference between group and phase velocity; (c) the elliptically polarised signal after rotation using Eq. (\ref{eq:rotation_between_columns}); (d) the signal in (a) after the polarisation correction or circularisation; (e) the circularly polarised signal after the  phase-shift given by the difference between group and phase velocity; (f) the circularly polarised signal after the rotation. Note that, for an elliptic polarisation, the phase shifted signal and the rotated signal are different, while for a circular polarisation, the phase shifted and the rotated signals are almost identical.}\label{fig:set_of_hodograms}
  \end{center}
\end{figure}

Eq. (\ref{eq:rotation_between_columns}) plays a key role as it links the phase shift $\Delta \phi$ caused by dispersion (e.g., see Eq. (\ref{eq:constant_phase_shift})) with a rotation of an angle $\Delta \phi$ in the $x$-$z$ plane. Fig. \ref{fig:set_of_hodograms} helps to understand that this correspondence is valid for circularly polarised signals only. The top-left panel (a) shows the elliptical hodogram (or particle motion plot) of the signal of Eq. (\ref{eq:stationary_phase_equations}) at the source position $x$=0, assuming $\Delta \omega$=$2\pi*0.5$ rad/s, $\omega_0$=$2\pi*5$ rad/s, $r_x$=$0.5$, and $r_z$=$1$. The same signal after the polarisation correction described in Eq. (\ref{eq:circularized_displacement}) is shown in the panel (d). Panels (b) and (e) show the two aforementioned signals after a phase-shift given by the difference between group and phase velocity as in Eq. (\ref{eq:constant_phase_shift}). Finally the panels (c) and (f) show the elliptically and circularly polarised signals after rotation using Eq. (\ref{eq:rotation_between_columns}). Note that, for an elliptic polarisation, the phase-shifted signal (b) and the rotated signal (c) are different, while for a circular polarisation, the phase shifted (e) and the rotated signals (f) are identical. 

\subsection{Quaternion representation}

Using the equivalence between vectors in 3D space and pure quaternions introduced in Section \ref{se:quat}, the displacement components in Eq. (\ref{eq:u_xm}) can be written in quaternion notation, as follows:

\begin{align} \label{eq:quaternion_circularized_displacement}
\mathbf{\mathring{u}}_m &= \mathtt{i} \mathbf{\underline{u}}_{x,m} + \mathtt{k} \mathbf{\underline{u}}_{z,m} , \qquad m=1,..., M,
\end{align}
where $\mathbf{\mathring{u}}_{m}$ are the columns of a matrix with pure quaternion entries, that we denote $\mathbf{\mathring{U}} \in \mathbb{H}^{N \times M}$
\begin{align} \label{eq:q_notation_matrix}
\mathbf{\mathring{U}} &= \mathtt{i} \mathbf{\underline{U}}_x + \mathtt{k} \mathbf{\underline{U}}_z 
\end{align}
describing the quasi-monochromatic Rayleigh wave recorded at the receiver positions. Such a matrix constitutes the ``quaternion brick'' that represents multi-component seismic data within a narrow frequency band around $\omega_0$ after the infinite group-velocity and the polarisation corrections.

As discussed in the previous section, a set of data that records a dispersive and circularly-polarised wave at equidistant locations can be seen as a collection of traces rotating by an angle $\Delta \phi$ in the $x$-$z$ plane. Recalling Eq. (\ref{eq:quaternion_rotation_rules}), this rotation can be written in a compact way using quaternions. 
Therefore, the quaternion brick in Eq. (\ref{eq:quaternion_circularized_displacement}) can be written as an outer product as follows:
% EQ 21
\begin{equation} \label{eq:q_rank1}
\mathbf{\mathring{U}} = \mathring{\mathbf{u}}_{1}
\begin{bmatrix}
1 & \mathring{r}_{\phi} & \mathring{r}^2_{\phi} & \dots & \mathring{r}^{M-1}_{\phi} \\
\end{bmatrix} ^{\ast},
\end{equation}
where $\mathring{\mathbf{u}}_{1}$ is the quaternion trace recorded at the first sensor, and
% EQ 22
\begin{equation} \label{eq:r_phi_rotation}
\mathring{r}_{\phi} = \pm \cos \phi + \mathtt{j} \sin \phi
\end{equation}
is a unit quaternion representing a rotation of $\phi$ radians about the $\mathtt{j}$ direction (the $y$ axis). The choice of the sign in the real part determines whether the rotation is clockwise or anti-clockwise (prograde or retrograde motion of the particle displacement). 

Using the definition of eigenimage of Eq. (\ref{eq:qsvd}) for $i$=1, and substituting
% EQ 26
\begin{align} \label{eq:21_qbrick_svd}
\sigma_1&= \mid \mathbf{\mathring{u}_{1}} \mid \nonumber \\
\mathbf{\mathring{w}_1} &= \frac{\mathbf{\mathring{u}_{1}}} {\mid \mathbf{\mathring{u}_{1}} \mid} \nonumber \\
\mathbf{\mathring{v}_1} &=
\begin{bmatrix}
1 & \mathring{r}_{\phi} & \mathring{r}^2_{\phi} & \dots & \mathring{r}^{M-1}_{\phi} \\
\end{bmatrix} ^T 
\end{align}
into Eq. (\ref{eq:q_rank1}), we get
%EQ 
\begin{align} \label{eq:21_qbrick_rank1}
\mathbf{\mathring{U}}= \sigma_1 \mathbf{\mathring{w}_1} \mathbf{\mathring{v}_1}^\dagger
\end{align}
This demonstrates that the quaternion brick is a rank-1 matrix. A similar proof has already been given by \cite{LeBihan04}. 

This compact and simple representation of the quasi-monochromatic Rayleigh-wave mode is possible because the quaternion brick is both trace-by-trace phase-shifted (as a consequence of dispersion) and instantaneously circularly polarised. Because of these two characteristics, the quaternion brick is a signal that rotates in the $\mathtt{i}$-$\mathtt{k}$ plane while it propagates in the in-line direction. Therefore the quaternion brick, which is constituted by the quasi-monochromatic and modified Rayleigh-wave mode, maps in the first eigenimage, while all other signals (e.g. random noise, body waves or other Rayleigh wave modes) spread into all eigenimages. This property can be exploited to extract the quaternion brick from the seismogram by truncating the QSVD of the seismic dataset to its first eigenimage. 

Note that in this special situation, where only two components are non-zero, the formulation can be simplified using complex numbers. In fact, in this specific case, Eq. \eqref{eq:21_qbrick_rank1} can be written as:

\begin{equation}
  \mathbf{\mathring{U}} = \mathbf{\underline{U}}_x + \mathtt{i} \mathbf{\underline{U}}_z 
  = \sigma_1 
  \begin{bmatrix}
    1 & e^{\mathtt{i} \phi} & e^{\mathtt{i} 2\phi} & \dots & e^{\mathtt{i} (M-1)\phi} \\ 
  \end{bmatrix} ^{\ast} \mathbf{w_1} 
\end{equation} 

\begin{figure}   
  \centering
  \includegraphics[scale=1]{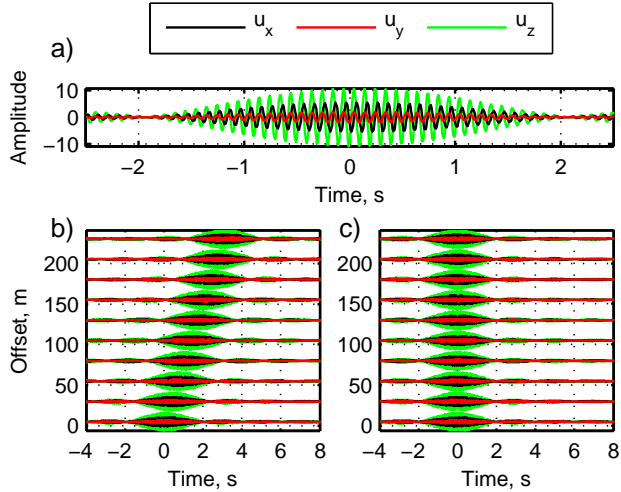} %l b r t 
  \caption{(a) Signal generated from Eq. (\ref{eq:stationary_phase_equations}) assuming $x$=0 (zero offset), $\omega_0$=2$\pi*10.25$ rad/s and $\Delta f$=0.25 Hz; the projection of the signal onto the three components is obtained rotating the direction of the multicomponent sensors by a 10-degree angle around the $x$ axis; $v_g$ and $v_{ph}$ are borrowed from the synthetic dataset described in the application example section; (b) the same signals evaluated at different offset to generate a synthetic seismogram; (c) traces corrected for group velocity.} \label{fig:5bis}
\end{figure}

\begin{figure}   
  \centering
  \includegraphics[width=\textwidth]{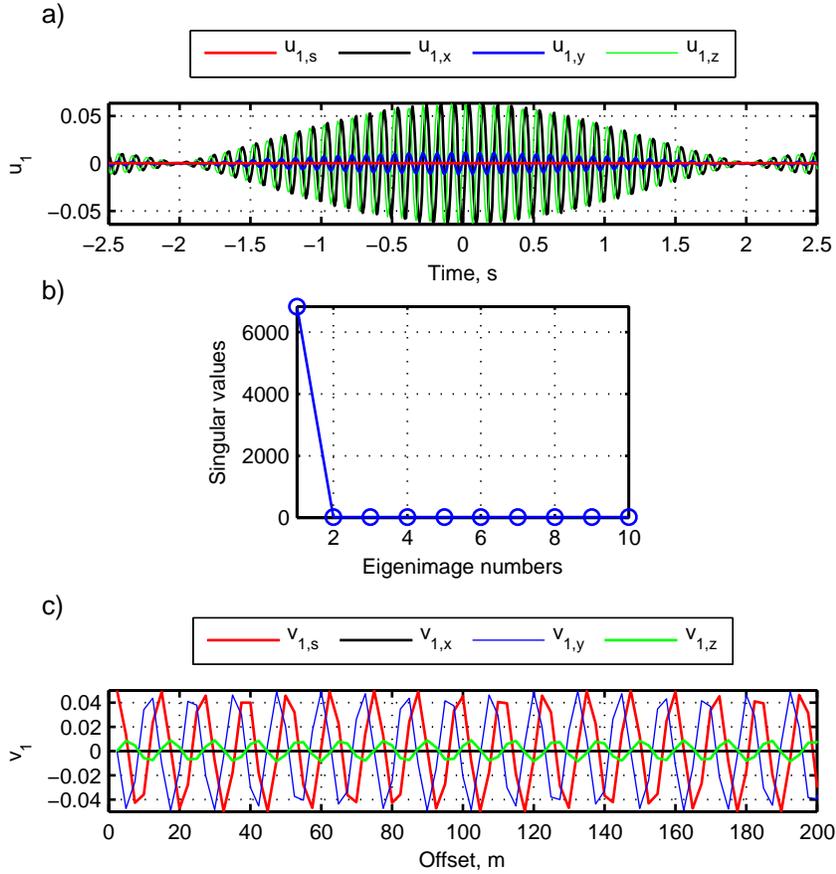} %l b r t 
  \caption{The results of quaternion SVD applied to the data of Fig. \ref{fig:5bis}(c), using $dx$=2.5 m, $x_{max}$=250 m. (a) The first column of the left singular vector $\mathring{\mathbf{w}}_1$; (b) the singular values; (c) the first column of the right singular value $\mathring{\mathbf{v}}_1$. Note that only one singular value is non-zero, as expected. Also note that $\mathring{\mathbf{w}}_1$ matches the first trace in Fig. \ref{fig:5bis}(a) after circularisation and that the $\mathring{\mathbf{v}}_1$ has three non-zero components that rotate as a sine ($z$ and $y$) and a cosine ($x$) in accordance with Eq. \eqref{eq:r_phi_rotation_general}.} \label{fig:5ter}
\end{figure}

However, the quaternion formulation allows us to treat also the more general situation in which the plane of motion does not match the plane defined by the sensor's $x$ and $z$ directions. The only condition, as required by Eq. \eqref{eq:quaternion_rotation_rules}, is that the particle motion is planar and the rotation axis is perpendicular to the plane of motion. Under these conditions, Eq. \eqref{eq:r_phi_rotation} can be rewritten in the general form:
\begin{equation} \label{eq:r_phi_rotation_general}
  \mathring{r}_{\phi} = \pm \cos \phi + \mathring{\omega} \sin \phi,
\end{equation}
Where $\mathring{\omega}$ is a pure unitary quaternion. Eq. \eqref{eq:21_qbrick_svd} and Eq. \eqref{eq:21_qbrick_rank1} still hold also for this more general case. The situation where the plane of motion of Rayleigh waves does not coincide with the $x$-$z$ plane may occur if the array of sensors has not been oriented correctly, i.e. the sensors are rotated with respect to the cartesian axis by a constant angle. An example of this situation is shown in Fig. \ref{fig:5bis}, in which we simulate a simple synthetic dataset by evaluating Eq. (\ref{eq:stationary_phase_equations}) at offsets ranging from 0 m to 250 m, with 2.5 m step between receivers, assuming $\omega_0$=2$\pi*10.25$ rad/s and $\Delta f$=0.25 Hz; the group velocity $v_g$ and phase velocity $v_{ph}$ are borrowed from the synthetic dataset described in the application example section. In this example, the directional sensors are mis-oriented, due to a 10 degree rotation around the in-line direction, such that the energy of the vertical direction is partially projected in the $y$ direction. The signal evaluated for offset zero ($x$=0) is shown in Fig. \ref{fig:5bis}(a); this trace is also evaluated at other offsets to generate the seismogram shown in Fig. \ref{fig:5bis}(b); the seismogram after the velocity and amplitude corrections is shown in Fig. \ref{fig:5bis}(c). The latter represents the quaternion brick for the frequency bandwidth around 10.25 Hz that is decomposed using QSVD. The first left singular vector $\boldsymbol{\mathring{w}_1}$, the real singular values $\sigma_i$ and the first right singular vector $\boldsymbol{\mathring{v}_1}$ are shown in Fig. \ref{fig:5ter}(a), (b) and (c), respectively. Note that only one singular value is non zero, as proved by Eq. (\ref{eq:21_qbrick_rank1}). It is also very interesting to observe that the vector $\boldsymbol{\mathring{w}_1}$ corresponds to the first trace of the seismogram (after the circularisation) and that $\boldsymbol{\mathring{v}_1}$ is a vector of quaternions representing the rotation by an angle $\phi$ as predicted by (\ref{eq:r_phi_rotation_general}). 

\subsection{The extraction method}

If we extend our analysis to a broad-band Rayleigh wave, we cannot approximate the group velocity and the phase velocity with a constant value. In fact, both group and phase velocities exhibit a frequency-dependent behavior which is properly described by the respective dispersion curves. Because of the broadband character of Rayleigh waves, adequately taking into account a wider frequency band is crucial in this theoretical analysis. To this end, we represent the displacement components of a broad-band Rayleigh-wave mode as a sum of narrow-band (or quasi-monochromatic) Rayleigh-wave modes with increasing central angular frequency $\omega_0$ and equal bandwidth $\Delta \omega$. In the formulation that we use, the phase- and group-velocities of the broad-band Rayleigh-wave mode are approximated as step functions of frequency, with constant values on each frequency interval $\Delta \omega$.
The separation into frequency bands is done by means of a filter bank with pass-bands $[\omega_j-\Delta \omega,\omega_j+\Delta \omega]$, for $j=1,2,...,J$. For each narrow frequency-band, the following steps are carried out:
\begin{enumerate}[1.]
\item Being known or measured the $v_g(\omega_j)$ of the mode, apply a time reduction $t'=t-x/v_g(\omega_j)$ to all displacement components such that the mode impinges at the same time at all offsets (infinite group-velocity correction);
\item Being known or measured the ellipticity value $c(\omega_j)$, apply the circularisation (see Eq. (\ref{eq:circularized_displacement})); before the correction, $c(\omega_j)$ is clipped to a finite range;
\item Apply QSVD to the data expressed in quaternion notation, and extract the first quaternion eigenimage (see Eq. (\ref{eq:q_rank1}) and Eq. (\ref{eq:r_phi_rotation})).
\end{enumerate}
Finally, the quaternion bricks are summed together to reconstruct the broad-band Rayleigh-wave mode.
The filter-bank parameters are chosen such that $\Delta \omega < \omega_1$, and $\omega_1- \Delta \omega$ and $\omega_J+\Delta \omega$ approximately equal the minimum and maximum frequencies at which the mode of interest propagates in the data. FIR filters are used to avoid distortion of the waveform in the filtering process. 

\subsection{Details on the first right singular vector: offset dependence}\label{se:det_1st_singular_vct}

Note from the outer product of Eq. (\ref{eq:21_qbrick_rank1}) that the offset dependence of the quaternion brick is all contained in the first right singular vector, $\mathring{\mathbf{v}}_1$. Furthermore, $\mathring{\mathbf{v}}_1$ is periodic. This has been shown graphically in Fig. \ref{fig:5ter}(c) and it can be also inferred from Eq. (\ref{eq:q_rank1}) and Eq. (\ref{eq:r_phi_rotation}).
The wavenumber $k$ of the sinusoids in Fig. \ref{fig:5ter}(c) is given by (see Eq. (\ref{eq:displacement_single_frequency}) and Eq. (\ref{eq:variables_for_displ_sing_freq})):
\begin{equation}
 k=\omega_0 \left( \frac{1}{v_{ph}} - \frac{1}{v_g} \right)
\end{equation}
This means that for each quasi-monochromatic mode centered in $\omega_0$ a single $k$ propagates. This is an approximation due to the finiteness of $\Delta \omega$ that nevertheless will also be verified in the next sections on a synthetic that we simulate using the reflectivity method. We notice that this periodicity of $\mathring{\mathbf{v}}_1$ can be exploited to infer the waveform of the Rayleigh wave mode for missing (not recorded) traces or in the case that the actual recorded waveform of the mode is disturbed by the interference with other signals, especially higher-order Rayleigh-wave modes. An example of this application will be shown in the next section.

\section{Tests on synthetic datasets}

We test this method on synthetic seismograms generated using OASES (Ocean Acoustics and Seismic Exploration Synthesis) \cite{Schmidt2004}, a modeling software based on the reflectivity method \cite{Kennett1974} that use a direct global matrix approach \cite{Ewing1957}. This software computes the entire seismic wavefield in flat, horizontally-stratified elastic media. Moreover, we simulate the Green's functions of the modes of the Rayleigh waves using the Matlab toolbox \emph{mat\_disperse} \cite{Lai1998}, which solves the eigenproblem of the Rayleigh waves by means of the modified R/T coefficients \cite{Chen1993}, a stabilised and optimised evolution of the generalised R/T coefficient method \cite{Kennett1974,Luco1983}. Note that \cite{Lai1998} does not simulate body waves. %\cite{Schmidt1988,Schmidt2004} \cite{FUCHS1968,Kennett1974} \cite{Schmidt1988, Ewing1957} \cite{Chen1993, Hisada1994,Lai1998}
 
To generate the synthetic seismograms we assume two flat and horizontally-stratified elastic models, whose layers are characterised by pressure- wave and shear-wave velocities ($v_p$ and $v_s$), density ($\rho$), and thickness ($H$) as described in Table \ref{tab:velocity_profile_1} (first model) and Table \ref{tab:velocity_profile_2} (second model). Both models are derived from a four-layer model proposed by \cite{Tokimatsu1992}. 
\begin{table} \centering
  \caption{First model used for testing our extraction procedure. For each layer: thickness, density, primary wave velocity, and shear wave velocity. The model is derived from \cite{Tokimatsu1992}).} \label{tab:velocity_profile_1}
  \: \: \: 
  \begin{tabu} to 0.9\textwidth {X[c]|X[c]|X[c]|X[c]|X[c]} \hline
    \textbf{Layer number} & \textbf{Thickness $H$ (m)} & \textbf{Density $\mathbf{\rho}$ (g/cm$^3$)} & \textbf{$\mathbf{v_p}$ (m/s)} & \textbf{$\mathbf{v_s}$ (m/s)}\\ \hline
    1 & 2 & 1.8 & 300 & 80 \\ 
    2 & - & 1.8 & 1000 & 120 \\ \hline
  \end{tabu} 
\end{table}
\begin{table} \centering
  \caption{Second model used for testing our extraction procedure. For each layer: thickness, density, primary wave velocity, and shear wave velocity. The model is derived from  \cite{Tokimatsu1992}.} \label{tab:velocity_profile_2}
  \: \: \: 
  \begin{tabu} to 0.9\textwidth {X[c]|X[c]|X[c]|X[c]|X[c]} \hline
    \textbf{Layer number} & \textbf{Thickness H (m)} & \textbf{Density $\mathbf{\rho}$ (g/cm$^3$)} & \textbf{$\mathbf{v_p}$ (m/s)} & \textbf{$\mathbf{v_s}$ (m/s)}\\ \hline
    1 & 2 & 1.8 & 300 & 80 \\ 
    2 & 4 & 1.8 & 1000 & 120 \\ 
    3 & 8 & 1.8 & 1400 & 180 \\ 
    4 & - & 1.8 & 1400 & 360 \\ \hline
  \end{tabu} 
\end{table}
The synthetic datasets are generated assuming a surface acquisition in which we propagate a source with cylindrical wavefront that simulates a 2D propagation. A Ricker-wavelet with 15 Hz peak frequency is used as source signature. The seismic wavefront is recorded with a 8 ms time-sampling rate by a 1D array of equally spaced receivers with 5 m receiver spacing and 250 m maximum offset.

\subsection{Simulating and extracting a single Rayleigh-wave mode: Proof of concept under optimal conditions} \label{se:appl-xmpl-single}

\begin{figure}   
  \centering
  \includegraphics[trim = 0mm 37mm 0mm 0mm, clip, scale=1]{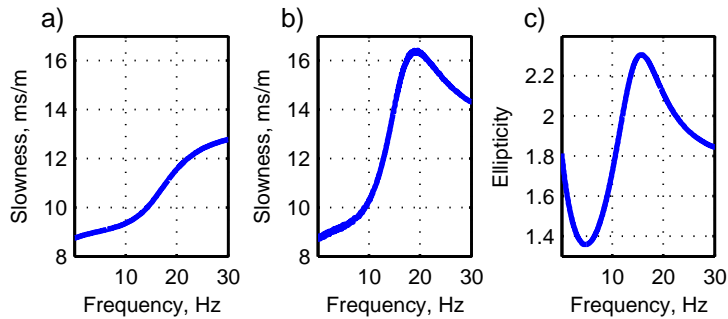} %l b r t 
  \caption{(a) Phase-slowness dispersion curve; (b) group-slowness dispersion curve; (c) ellipticity, i.e., ratio between the vertical and the horizontal displacement components. All the curves are referred to the mode of the Rayleigh wave propagating in the model described in Table \ref{tab:velocity_profile_2}} \label{fig:T-1}
\end{figure}

We start our analysis on the first model (Table \ref{tab:velocity_profile_1}), which is constituted by a thin layer over a faster half-space. In this simple model, a single Rayleigh-wave mode propagates (the fundamental mode) up to 21.5 Hz, which is the cutoff frequency of the first higher mode. Because we energise a frequency band centered on 15 Hz, the fundamental mode carries the most part of the energy in our tests. For this reason we can reduce the study of Rayleigh waves to the sole fundamental mode for this model. The dispersion characteristics of this mode are depicted in the panels (a) and (b) of Fig. \ref{fig:T-1} showing the phase-slowness and group-slowness dispersion curves, respectively. Note from Fig. \ref{fig:T-1} that this mode has normal dispersion, as expected. Fig. \ref{fig:T-1}(c) displays the ellipticity, i.e. the ratio between the two axes of the ellipse of polarisation of the Rayleigh-wave mode. In practice, the ellipticity is computed as the ratio in the frequency domain between the vertical and the horizontal components of the Green's function. Note that this ratio is always bigger than 1, meaning that the major axis of the ellipse of polarisation coincides with the vertical direction, and that the oscillations of these values with the frequency are within a small range.

As a proof of concept, we consider an optimal case in which the input dataset is constituted by a single Rayleigh-wave mode and no body waves nor Love waves are modelled. This test is set to validate the theoretical assumption described in Section \ref{se:Q_extraction}. The narrow-band analysis is done using a filter-bank spanning the range from 2.5 Hz to 28 Hz and FIR band-pass filters with 0.5 Hz bandwidth.
For this test, we use the modified R/T coefficient method \cite{Lai1998} to derive the displacements of the Rayleigh-wave mode that propagates in the two-layer model of Table \ref{tab:velocity_profile_1}, and we project the signal onto the three components by rotating the sensors’ system by 10 degrees around the $x$ axis. 

\begin{figure}   
  \centering
  \includegraphics[width=1\textwidth]{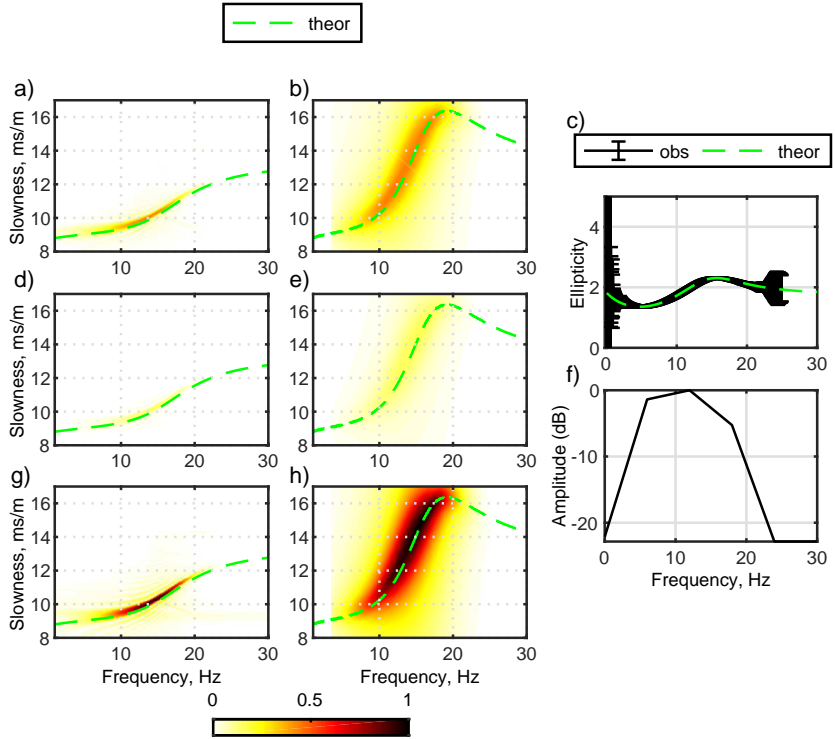} %l b r t 
  \caption{(a,d,g) Frequency-phase slowness transforms for $x$, $y$, and $z$ components with the theoretical frequency-phase slowness curve overlain (green dashed line); (b,e,h) frequency-group slowness transforms for $x$, $y$, and $z$ components with the theoretical curve overlain (green dashed line); (c) estimated ellipticity (black curve) with the theoretical ratio overlain (green dashed line); (f) amplitude spectrum of the input wavelet.} \label{fig:T-3}
\end{figure}

The extraction method that we propose relies on the knowledge of the group-slowness dispersion curve and of the frequency-dependent ellipticity of the Rayleigh-wave mode. This information is used respectively to perform the time-shift correction and the circularisation. In this paper, as anticipated in Section \ref{se:Q_extraction}, we use the theoretical values for both group slowness and ellipticity as shown in Fig. \ref{fig:T-1}(b), and (c). Nevertheless these parameters can be estimated from the data, computing the frequency-group slowness transform (a slant stack of the envelopes for each narrow band of the data) and the ratio, in the frequency domain, between the vertical and horizontal displacement components. A comparison between the estimated dispersion and polarisation parameter and the theoretical curves is shown in Fig. \ref{fig:T-3}. The panels (a), (d) and (g) show the f-phase slowness (f-p) transforms of the $x$, $y$, and $z$ components of the displacement. The f-p transform consists in a double linear transformation: a slant stack that transforms the data to the time intercept-phase slowness ($\tau$-p) domain followed by a 1D Fourier transform that maps the data in the frequency-phase slowness (f-p) domain \cite{McMechan1981}. The resulting images represent an accurate estimate of the phase-slowness dispersion curve that closely matches the theoretical curve (dashed green line); the panels (b), (e), and (h) display the frequency-group slowness transforms that provide a good estimate of the theoretical curve, although with less resolution than the phase-slowness curve; the panel (c) shows the ellipticity estimated from the seismograms (black line). Note the good match with the theoretical value (dashed green line) for all frequencies except at the low and high frequency bounds, as expected by observing the frequency spectrum of the Ricker wavelet used as input (f).% and the and finally, the panel (f) shows the amplitude spectrum of the Ricker wavelet used to generate the synthetic seismograms.

The input synthetic seismograms for the $x$, $y$, and $z$ components are shown in Fig. \ref{fig:T-4}(a), (d), and (g), respectively.
The remaining panels of Fig. \ref{fig:T-4} show the results of the extraction procedure for the single-mode Rayleigh wave: the panels (b), (e), and (h) are the $x$, $y$, and $z$ extracted datasets; (c), (f) and (i) are the residuals, i.e., the differences between (a) and (b), (d) and (e), and (g) and (h), respectively. Note that the Rayleigh-wave mode has been almost completely recovered. In fact, the residual energy is only a small fraction of the input energy: 2.4$\%$ for the $x$ component, 3.6$\%$ for the $y$ component, and 2.4$\%$ for the $z$ component.

\begin{figure}   
  \centering
  \includegraphics[width=1\textwidth]{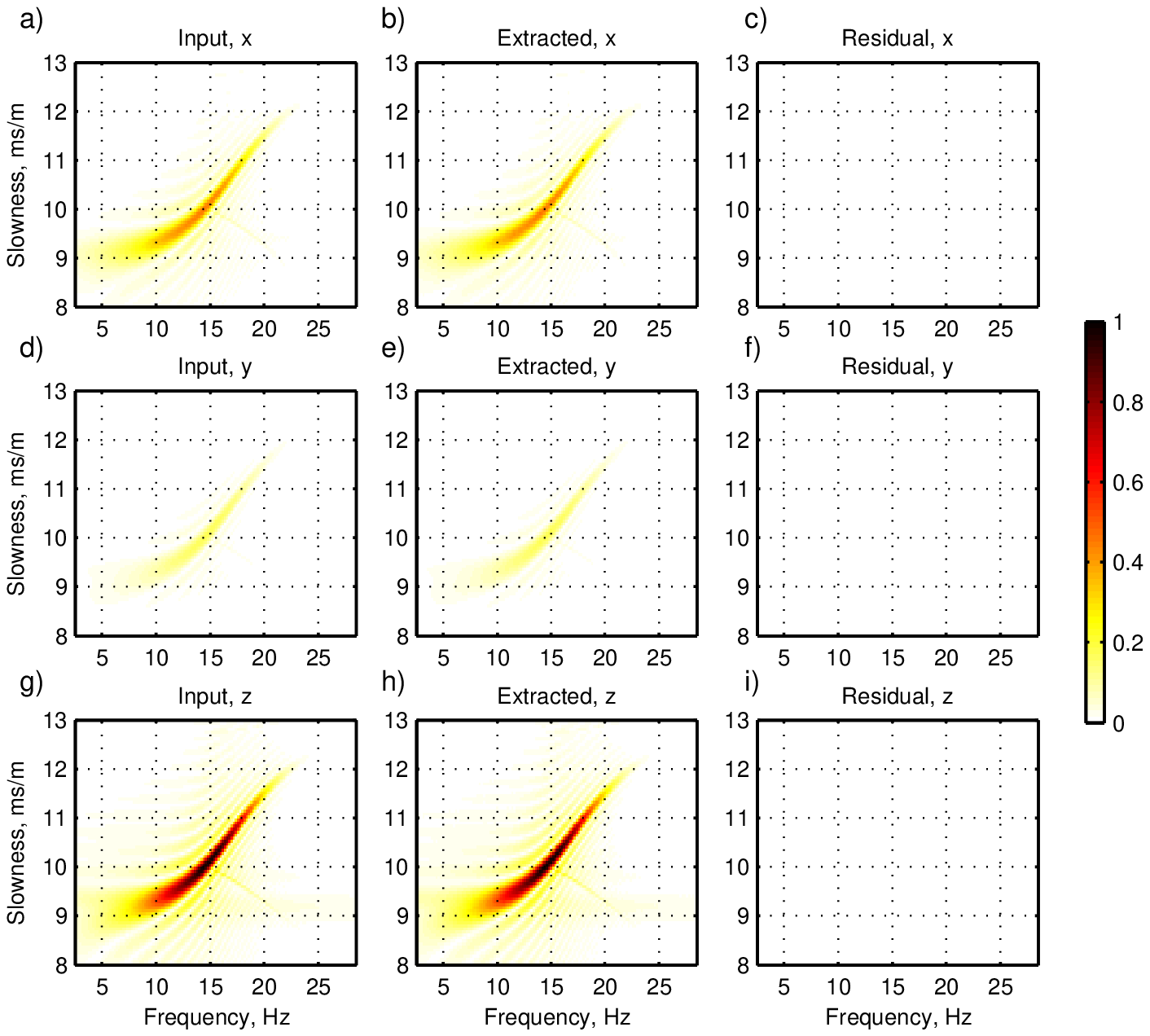} %l b r t 
  \caption{(a) The input synthetic dataset in the f-p domain, $x$ component; (b) the f-p transform of the corresponding extracted data; (c) the f-p transform of the residual (a) minus (b); (d) the input synthetic dataset in the f-p domain, $y$ component; (e) the f-p transform of the corresponding extracted Rayleigh wave; (f) the f-p transform of the residual (d) minus (e); (g) the input synthetic dataset in the f-p domain, $z$ component; (h) the f-p transform of the corresponding extracted data; (i) the f-p transform of the residual (g) minus (h). Note that the dispersive characteristic of the Rayleigh wave is entirely represented in the extracted panels for all components.} \label{fig:T-5}
\end{figure}

We can evaluate the quality of the extraction also by comparing the data in the f-p domain. Fig. \ref{fig:T-5} shows the input, the extracted and the residual datasets in the f-p domain, using the same panel-scheme of Fig. \ref{fig:T-4}. Note that almost no dispersive signal is visible in the residual panels (Fig. \ref{fig:T-5}(c), (f), and (i)). When computed in the f-p domain, the ratio between the residual energy and the input energy is 2.2$\%$, 2.7$\%$ and 1.9$\%$ for the $x$, $y$ and $z$ components, respectively.
These results validate the theoretical assumption that it is sufficient to extract the first quaternion eigenimage, for each frequency of the signal, to retrieve the whole energy of a broad-band Rayleigh-wave mode.

\begin{figure}   
  \centering
  \includegraphics[width=0.95\textwidth]{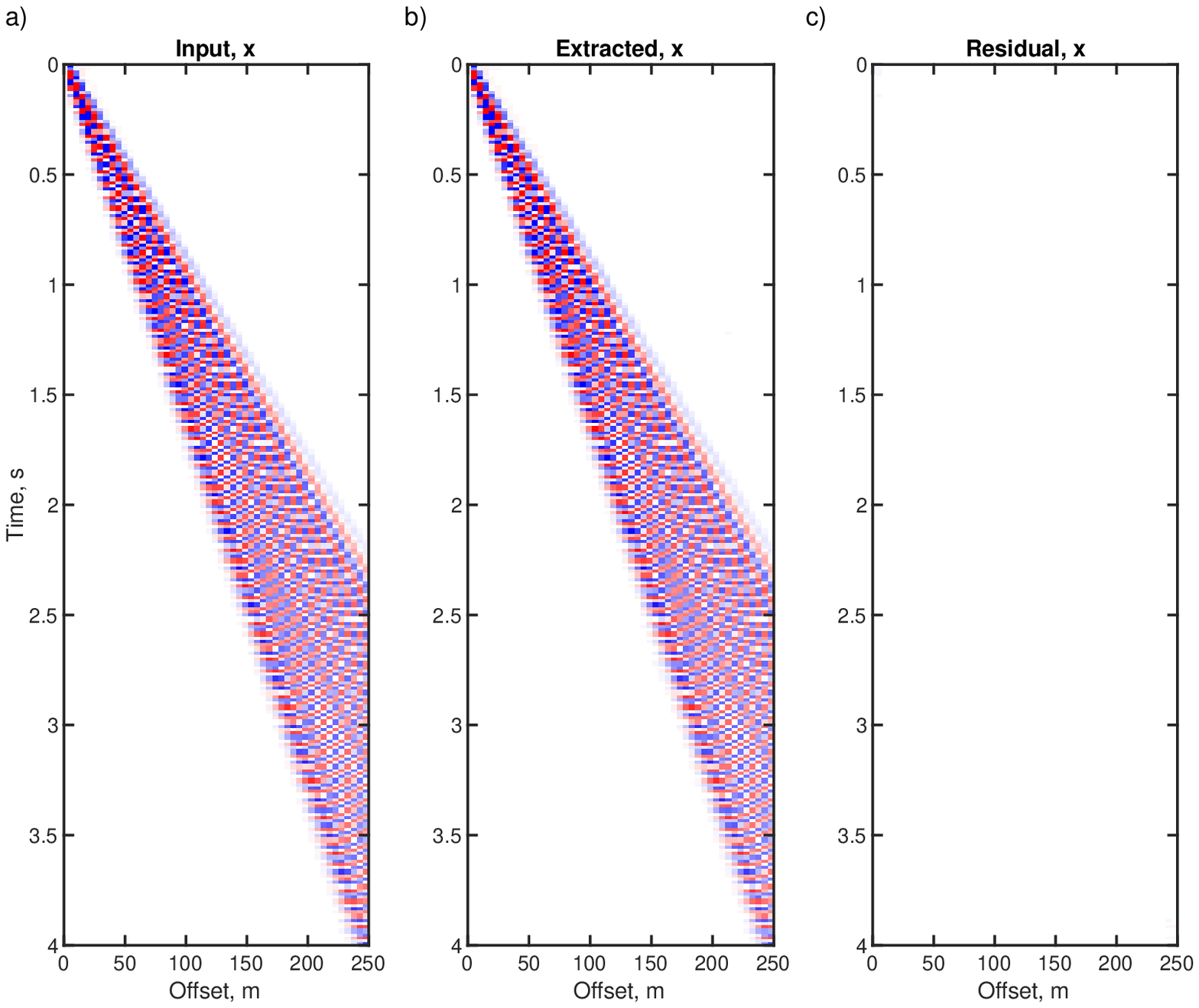} \\ %l b r t 
  \includegraphics[width=0.95\textwidth]{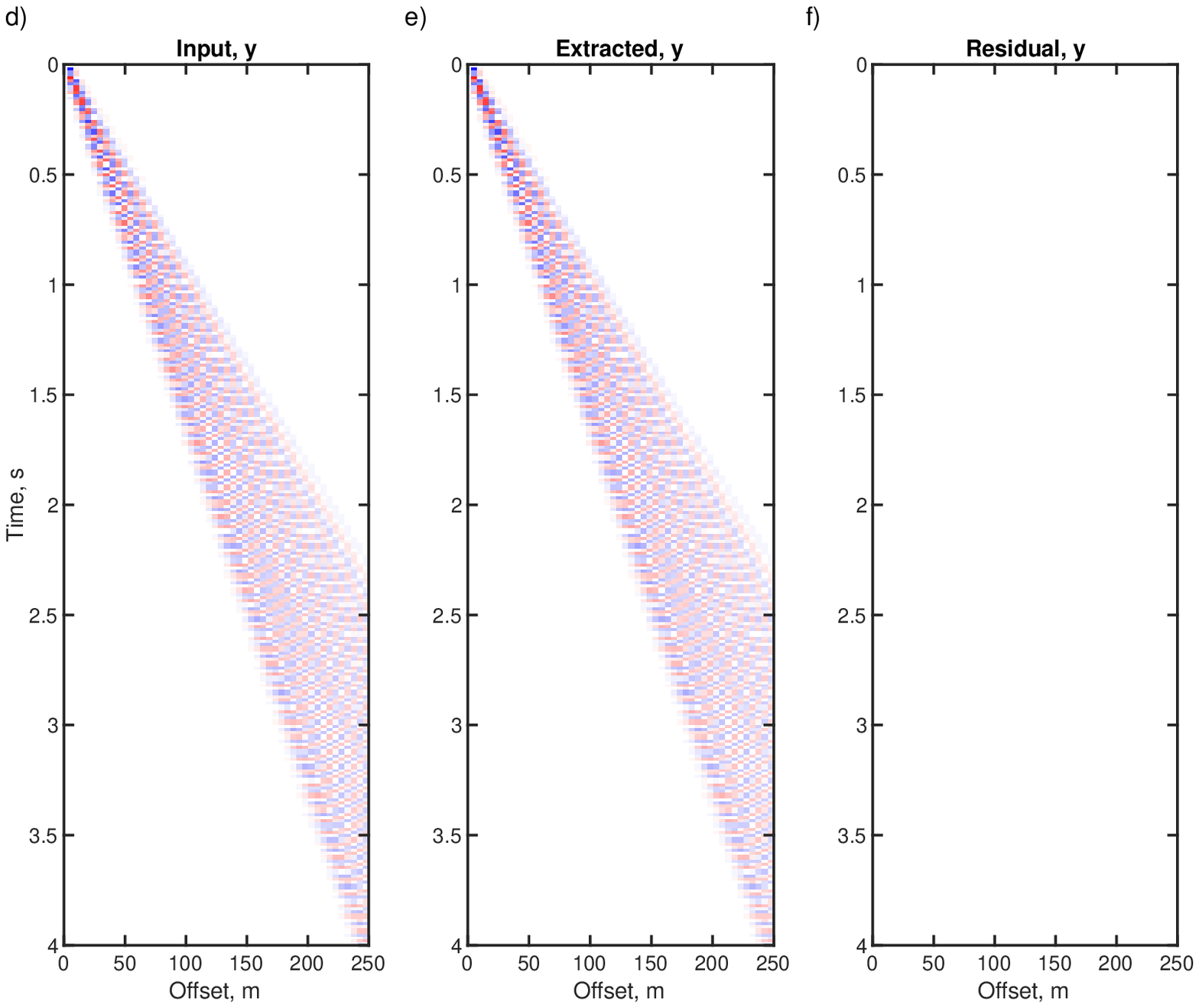} %l b r t 
  \caption{Synthetic dataset containing the Rayleigh wave, simulated using the modified R/T coefficients method \cite{Lai1998}. (a) the input synthetic dataset: single-mode Rayleigh wave, $x$ component; (b) the corresponding extracted Rayleigh wave; (c) the residual i.e. the difference between (a) and (b); (d) the input synthetic dataset: single-mode Rayleigh wave, $y$ component; (e) the corresponding extracted Rayleigh wave; (f) the residual i.e. the difference between (d) and (e). Note that the Rayleigh wave is almost entirely contained in the extracted panels for both components.} 
\end{figure}

\begin{figure}   \ContinuedFloat
  \centering
  \includegraphics[width=0.95\textwidth]{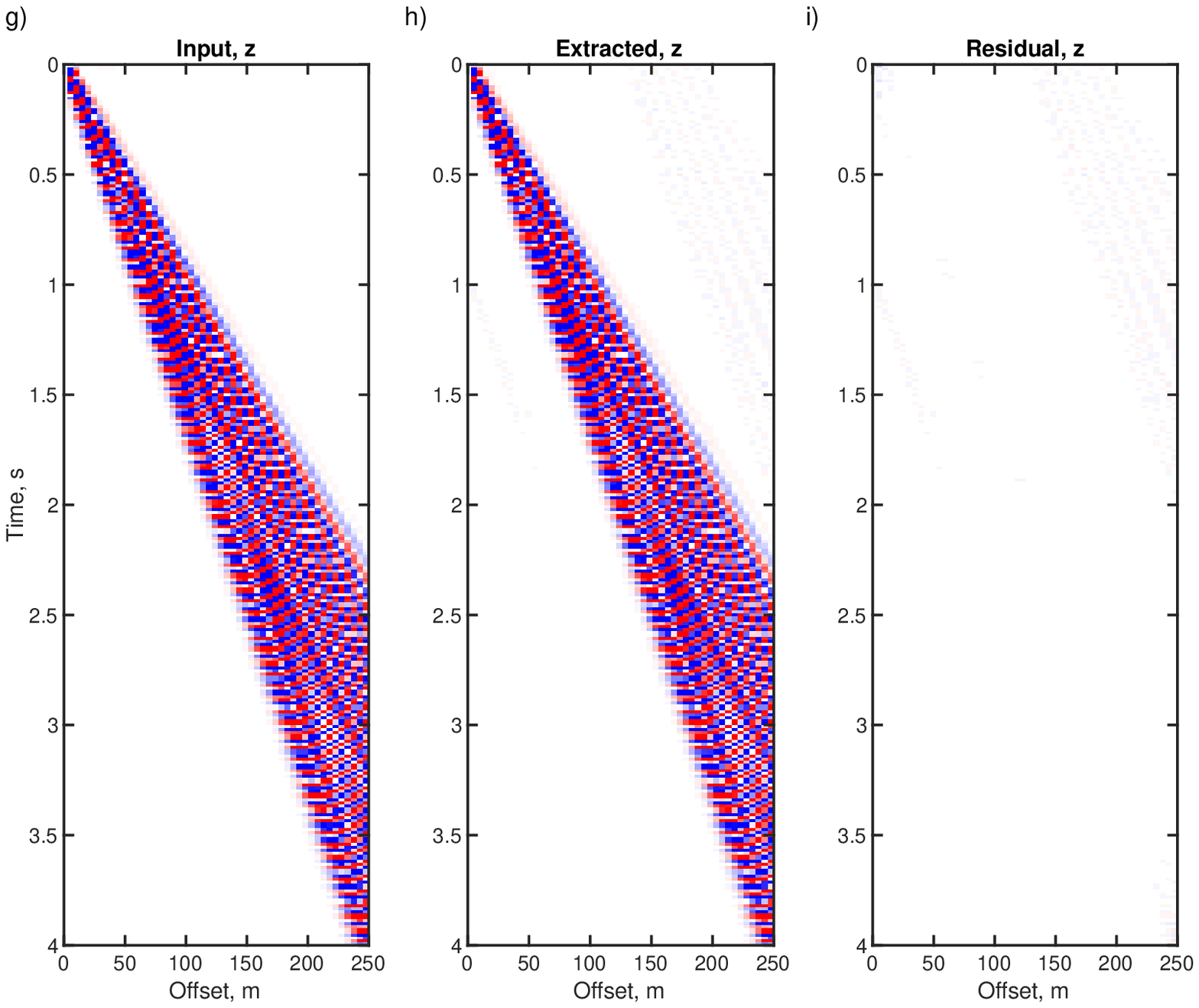} %l b r t 
  \caption{(Continued) Synthetic dataset containing the Rayleigh wave, simulated using the modified R/T coefficients method \cite{Lai1998}. (g) the input synthetic dataset: single-mode Rayleigh wave, $z$ component; (h) the corresponding extracted Rayleigh wave; (i) the residual i.e. the difference between (g) and (h). Note that the Rayleigh wave is almost entirely contained in the extracted panels.} \label{fig:T-4}
\end{figure}

\subsection{Separating a single-mode Rayleigh wave from body waves}

In this test, we use the software OASES to solve for the Green's function that includes both body waves and Rayleigh waves. Differently from the previous test, we do not rotate the sensors' system and hence the displacements of the wavefield are on the $x$ and $z$ components only. However, elastic model, source wavelet, and source/receivers positions are the same as in the previous test. Consequently, we can use the same parameters to perform the velocity correction, and polarisation correction. In addition, we use the same decomposition in narrow bands that we employed in the previous test.

\begin{figure}   
  \centering
  \includegraphics[width=0.95\textwidth]{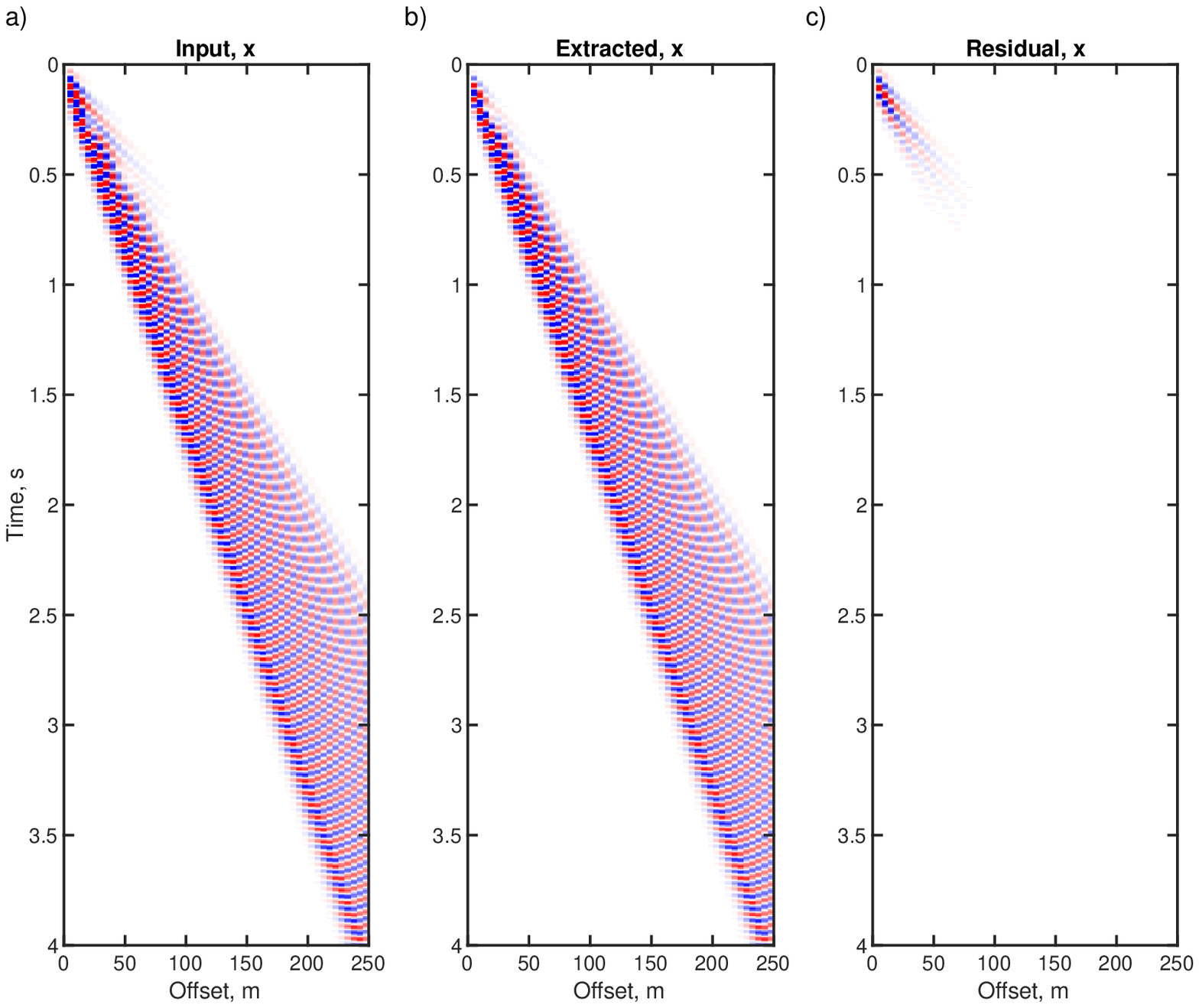} \\%l b r t 
  \includegraphics[width=0.95\textwidth]{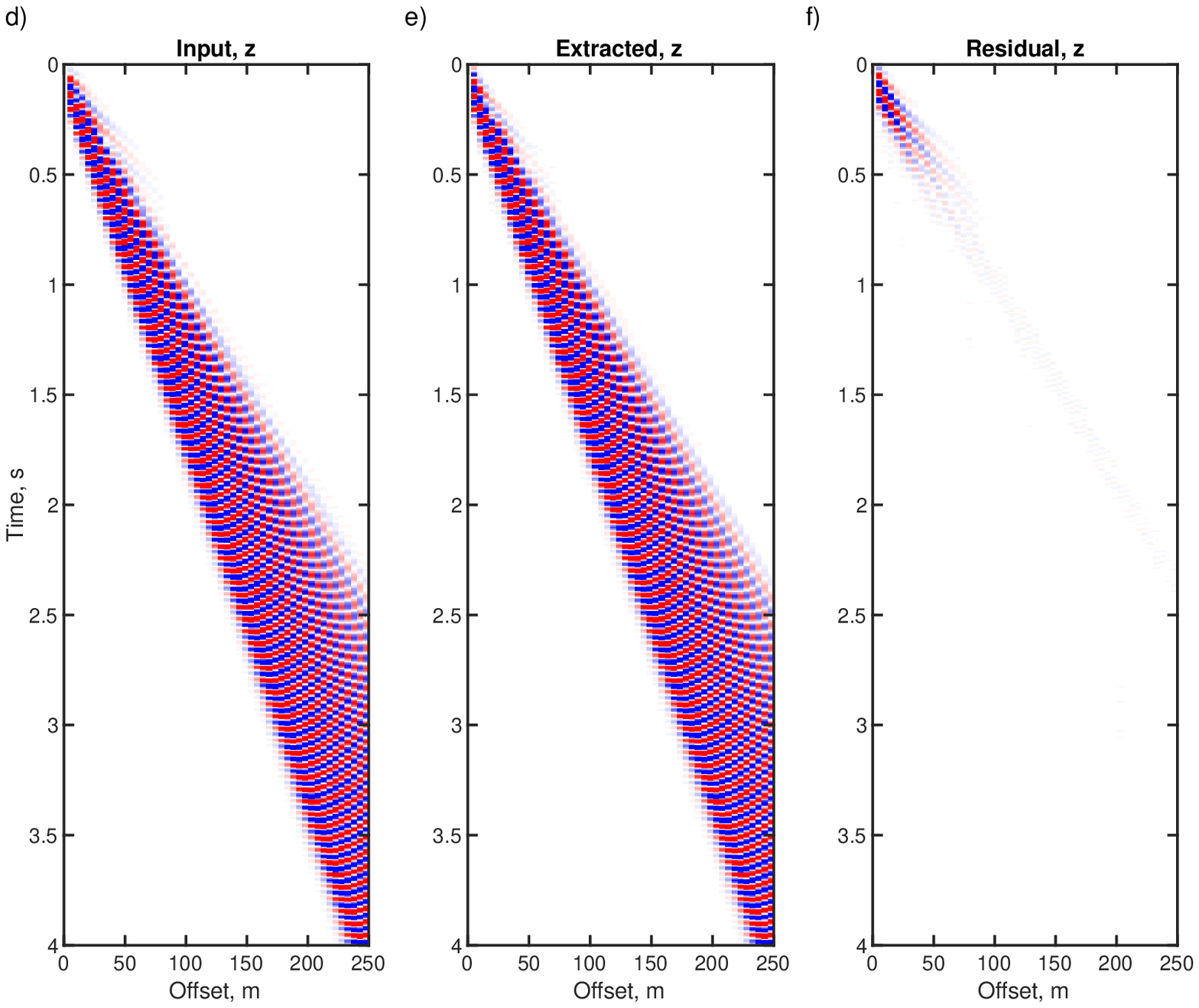}
  \caption{Synthetic dataset containing both Rayleigh-wave and body waves, simulated with the reflectivity method \cite{Kennett1974}. (a) the input dataset, $x$ direction; (b) the corresponding extracted dataset; (c) the residual, i.e., the difference between (a) and (b). (d) the input dataset, $z$ direction; (e) the corresponding extracted dataset; (f) the residual, i.e., the difference between (d) and (e). Note that the residual mainly contains high-velocity body waves with a small residual of the dispersive single-mode Rayleigh wave.} \label{fig:T-6}
\end{figure}

The results of the extraction are shown in Fig. \ref{fig:T-6} for both the $x$ and $z$ components. The panels (a) and (d) show the input components, (b) and (e) show the extracted dataset, and (c) and (f) show the residual data, i.e., the difference between the input and the extracted data. 
Note that the low-velocity dispersive train-wave is almost entirely contained in the extracted data, while almost no energy of the Rayleigh wave is present in the residual panels, indicating that the surface wave has been correctly extracted. Conversely, the higher-velocity body waves are clearly visible in the residual panel. 

\begin{figure}   
  \centering
  \includegraphics[width=1\textwidth]{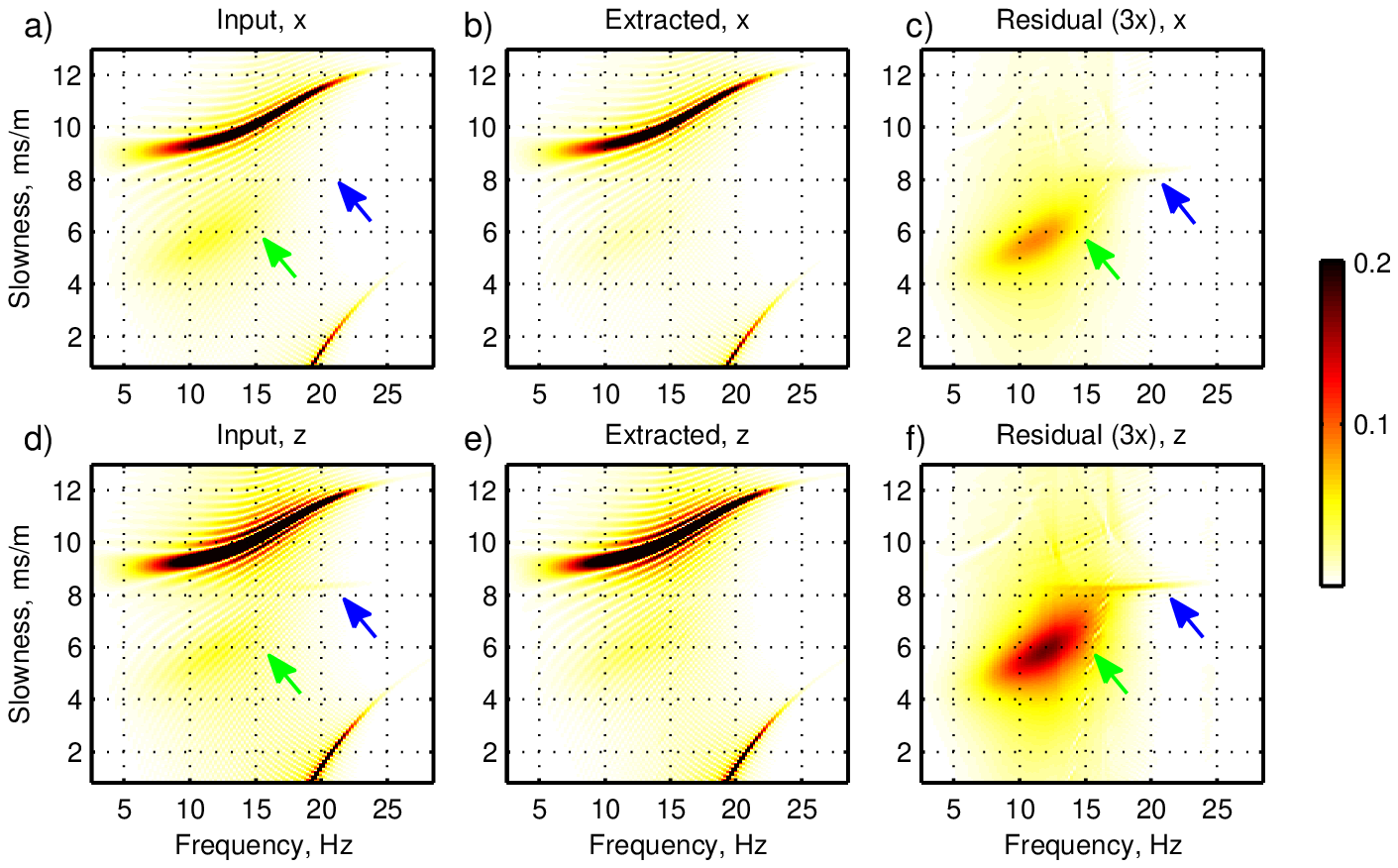} %l b r t 
  \caption{(a) The input synthetic dataset in the f-p domain, $x$ component; (b) the f-p transform of the corresponding extracted data; (c) the f-p transform of the residual (a) minus (b), scaled by a factor of three; (d) the input synthetic dataset in the f-p domain, $z$ component; (e) the f-p transform of the corresponding extracted Rayleigh wave; (f) the f-p transform of the residual (d) minus (e), scaled by a factor of three. Note that the dispersive characteristics of the Rayleigh wave is almost entirely represented in the extracted panels for both components. On the contrary, the body wave at 8 ms/m (blue arrow) and the blob (red arrow) are mainly visible in the residuals.} \label{fig:T-8}
\end{figure}

The results of the extraction can also be analysed in the f-p domain shown in Fig. \ref{fig:T-8}. The input panels (a) and (d) are dominated by the fundamental mode of the Rayleigh wave. In fact, only a hint of body wave is visible in the panel showing the $z$ component (d). This event, possibly a converted wave, shows up as a quasi-horizontal event with slowness equal to 8 ms/m (blue arrow). The other events are: a dispersive energy blob near slowness 6 ms/m that we have not classified (probably due to a dispersive guided wave, an artifact of the f-p transform, or another event, see green arrow) and an artefact, due to aliasing, in the bottom-right corner also evident in the $x$-$t$ domain. Note that the dispersive Rayleigh-wave mode, is correctly mapped into the extracted panels (b) and (e), with little residual visible on panels (c) and (f). This denotes that the dispersion characteristics have been almost entirely recovered with the extracted procedure. Also note that the residual panels contains almost entirely the converted body wave near $p$ = 8 ms/m and the energy blob near $p$ = 6 ms/m.

\subsection{Separating the fundamental mode from the first higher mode}

\begin{figure}   
  \centering
  \includegraphics[trim = 0mm 40mm 0mm 0mm, clip, width=\textwidth]{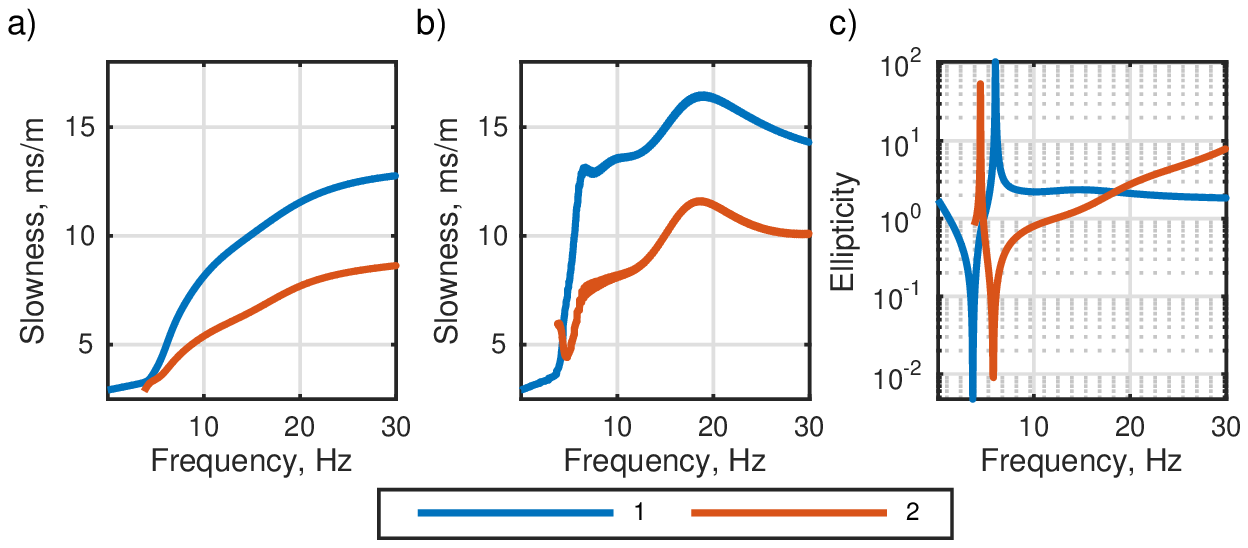} %l b r t 
  \caption{(a) Phase-slowness dispersion curves, (b) group-slowness dispersion curves, and (c) ellipticity for the first two modes of propagation of Rayleigh waves in the ground model of Table \ref{tab:velocity_profile_2}.} \label{fig:T-9}
\end{figure}

In this section we test our extraction algorithm on a synthetic generated using the model described in Table \ref{tab:velocity_profile_2}. For this model, which is constituted by three layers with increasing $v_p$ and $v_s$, over a stiffer half-space, multiple Rayleigh-wave modes propagate in the frequency band of interest (approximately 4-26 Hz). The theoretical phase-slowness and group-slowness dispersion curves of the first two modes of propagation for this model are shown in Fig. \ref{fig:T-9}(a) and (b), respectively.
Fig. \ref{fig:T-9}(c) displays the ellipticity of the first two modes. Note that at approximately 4 Hz this value tends toward 0, which corresponds to a linear polarisation in the horizontal direction, while at approximately 6.5 Hz the ellipticity tends toward infinity, which corresponds to a linear polarisation in the vertical direction. At these frequencies, the particle motion direction changes from retrograde to prograde ($\sim$4 Hz) and from prograde to retrograde ($\sim$6.5 Hz). Note also the high variability of the ellipticity in the low-frequency range ($<$10 Hz), while for higher frequencies ($>$10 Hz) it becomes much more stable.

We illustrate the results of the extraction procedure on a synthetic dataset generated using the R/T coefficient method \cite{Lai1998}. For simplicity, only the first two modes of the Rayleigh wave, i.e. the fundamental and the first higher mode, are modelled, and we do not rotate the sensors' system. 

\begin{figure}   
  \centering
  \includegraphics[width=1\textwidth]{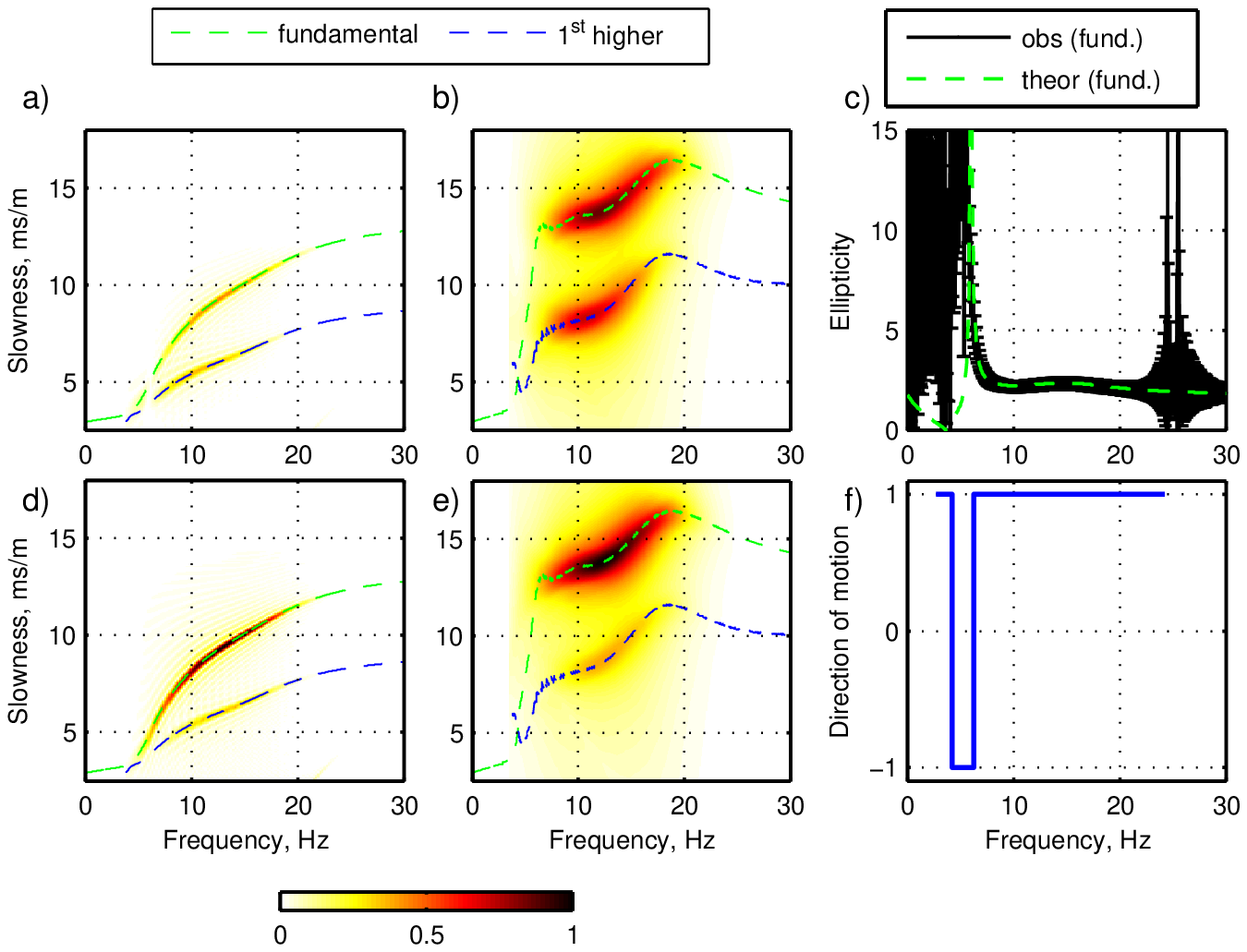} \\
  \caption{(a) frequency-phase slowness transform for the $x$ component with the theoretical curves overlain (dashed line); (b) frequency-group slowness transform for the $x$ component with the theoretical curves overlain (dashed line); (c) frequency-phase slowness transform for the $z$ component with the theoretical curves frequency-phase slowness overlain (dashed line); (d) frequency-group slowness transform for the $z$ component with the theoretical curves overlain (dashed line); (e) estimated ellipticity (black curve) with the theoretical value overlain (green dashed line); the sign of the cosine of Eq. \eqref{eq:r_phi_rotation_general} representing the direction of the particle motion ($-1$: prograde motion; $+1$: retrograde motion).} \label{fig:T-12}
\end{figure}

The two dispersive wave trains can be clearly recognised in both f-p domain and in the frequency-group slowness domain. The f-p transforms for the $x$ and $z$ components are shown in Fig. \ref{fig:T-12}(a) and (c), respectively, with the theoretical curve of the fundamental and first higher modes overlain (green and blue dashed line). In this domain the two modes are clearly separated apart from a frequency interval around approximately 5 Hz in which they are superimposed. The frequency-group slowness transforms for the $x$ and the $z$ components, are shown in Fig. \ref{fig:T-12}(b) and (d), respectively, with the corresponding theoretical curve for the first two modes also shown with a dashed line. The energy of the two modes is similar for the $x$ component, whereas for the $z$ component the fundamental mode is more energetic. This justifies our strategy of choosing to extract the fundamental mode. Fig. \ref{fig:T-12}(e) shows the ellipticity computed on a portion of data containing mainly the fundamental mode. The estimated ellipticity (black line) matches well the theoretical value (green dotted line), apart from the low frequency range where the higher mode might interfere. Fig. \ref{fig:T-12}(f) shows the sign of the cosine of Eq. \eqref{eq:r_phi_rotation_general} that represents the direction of the particle motion. Note the shift from retrograde motion (positive value) to prograde (negative value) at 4 Hz, and again to retrograde motion at 6.5 Hz.

\begin{figure}   
  \centering
  \includegraphics[width=1\textwidth]{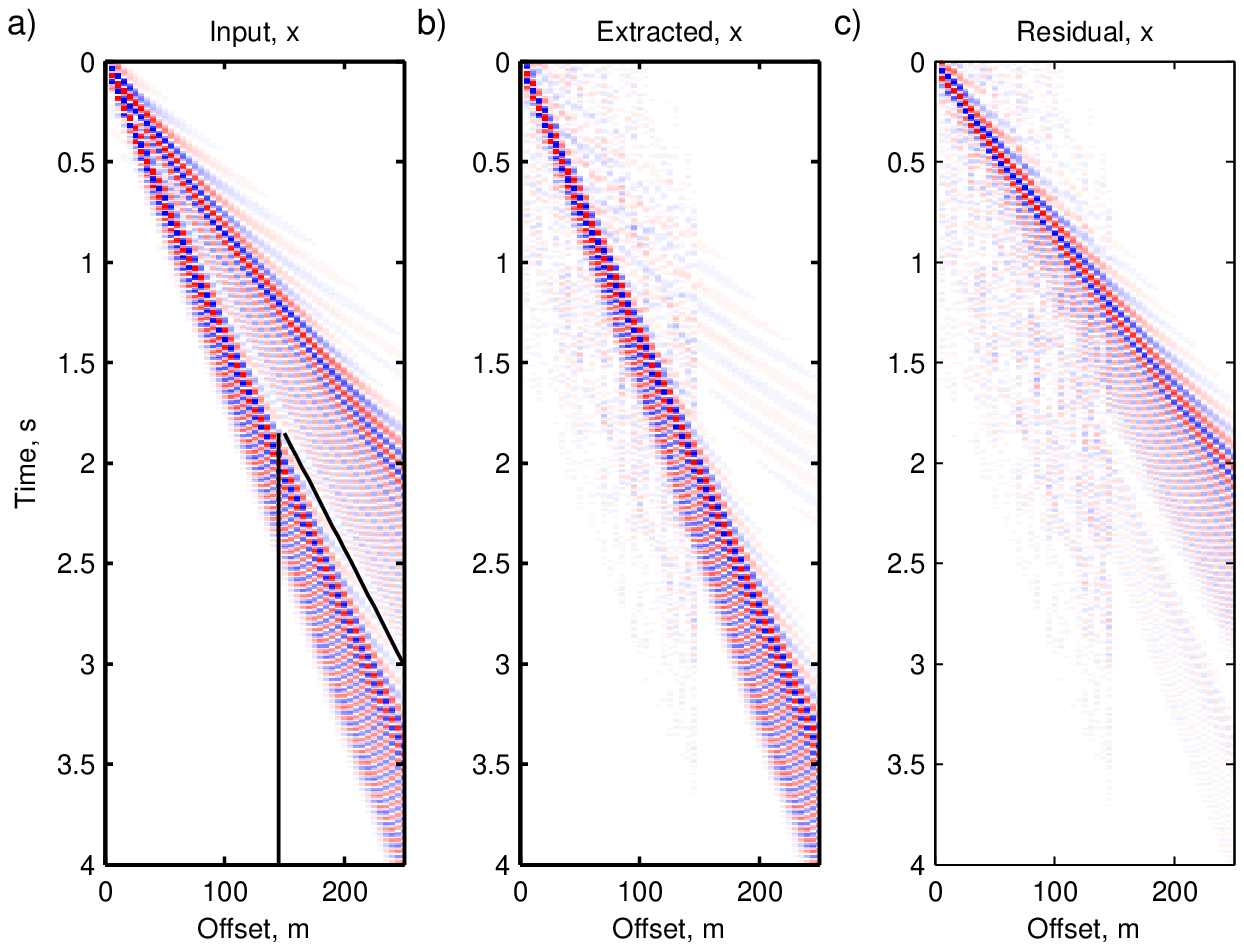} \\%l b r t 
  \includegraphics[width=1\textwidth]{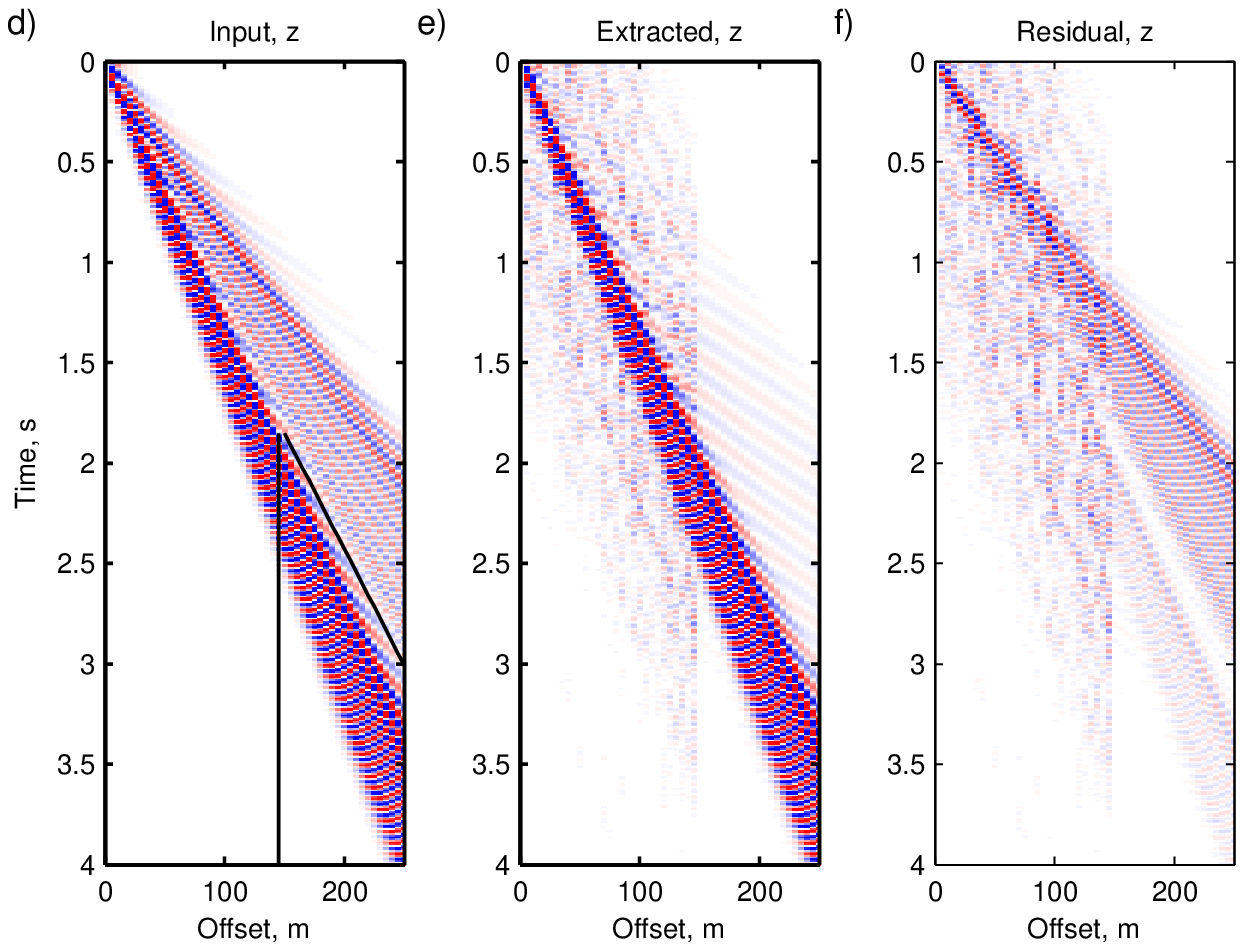} \\
\caption{(a) the input synthetic dataset, $x$ direction; the black lines delimit the portion containing the fundamental mode only; (b) extracted fundamental mode, in which the near offsets have been extrapolated using QSVD, $x$ direction; (c) residual, that is the difference between the input dataset (a) containing the two modes and (b). (d) the input synthetic dataset, $z$ direction; the black lines delimit the portion containing the fundamental mode only; (e) extracted fundamental mode, in which the near offsets have been extrapolated using QSVD, $z$ direction; (f) residual, that is the difference between the input dataset (d) containing the two modes and (e).} \label{fig:T-16}
\end{figure}

\begin{figure}   
  \centering
  \includegraphics[width=1\textwidth]{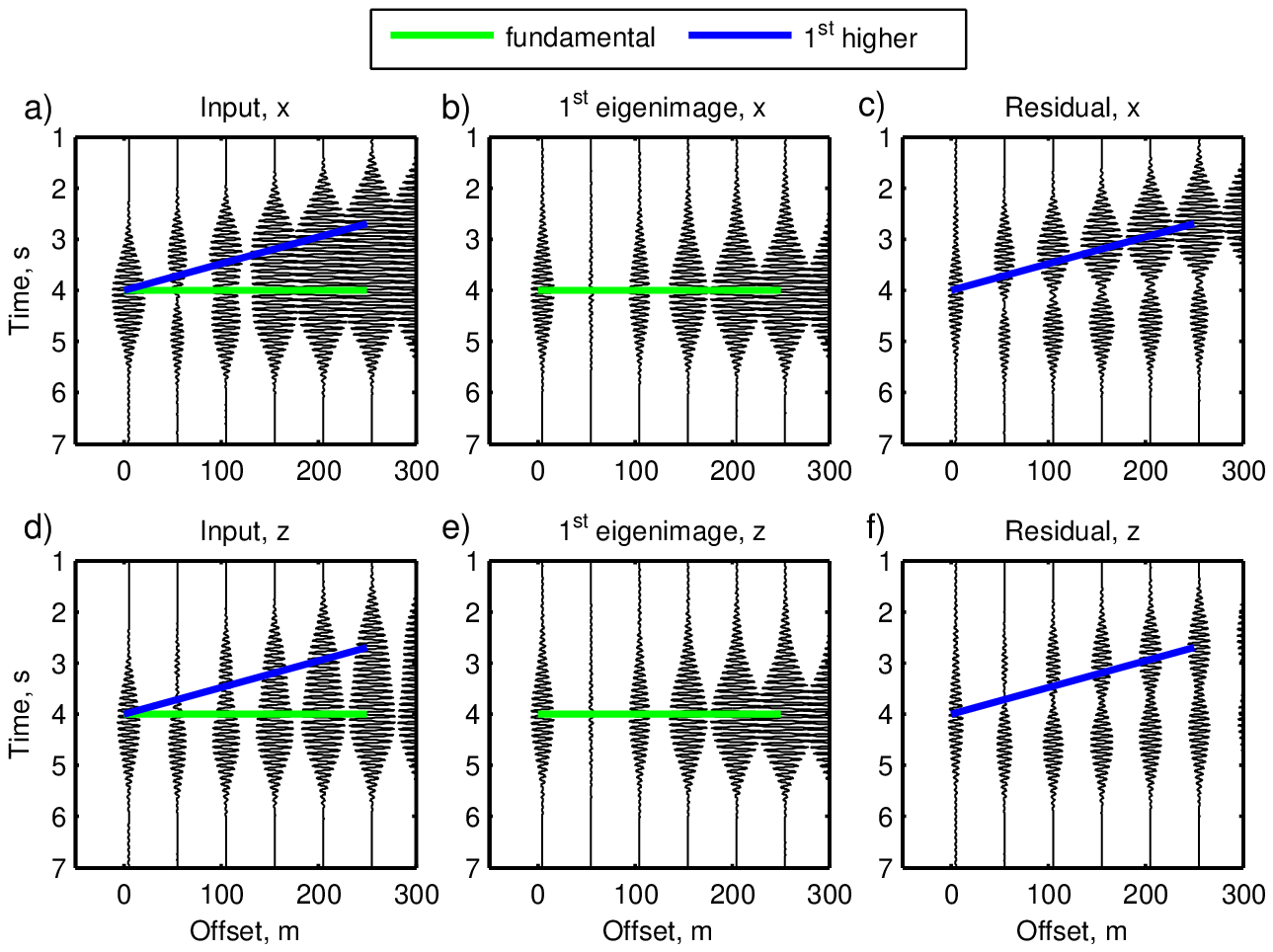} \\
  \caption{(a) the $x$ component of the 8 Hz quaternion brick before the application of QSVD; (b) the $x$ component of the first eigenimage; (c) the sum of the residual eigenimages ($x$ component); (d) the $z$ component of the 8 Hz quaternion brick before the application of QSVD; (e) the $z$ component of the first eigenimage; (f) the sum of the residual eigenimages ($z$ component); the dashed lines represent the expected arrival times of the wave packets of the fundamental (green) and first-higher (blue) Rayleigh-wave modes after the group-velocity correction; note that the amplitude of the extracted Raileigh-wave mode is low at the second trace, suggesting that the mode is not well reconstructed.} \label{fig:T-15bis}
\end{figure}

The $x$ and $z$ components of the input data are shown in Fig. \ref{fig:T-16}(a) and (d). 
The dispersive wave-trains of the two modes are clearly visible in the time-offset domain, interfering in the near-offset portion and progressively separating at larger offsets due to the different velocities of propagation. Although not shown here for the sake of brevity, in this scenario, the extraction procedure fails to correctly distinguish and separate the two signals, leaving part of the first-higher mode in the extracted data-sets that should only contain the fundamental mode. We explain this behavior with the difficulty of the QSVD to reject energetic signals, as their contribution to the first eigenimage may not be negligible. Fig. \ref{fig:T-15bis} shows this behaviour for one quaternion brick (8 Hz) comparing the modified signals before QSVD (panels (a) and (d)), the first eigenimages (panels (b) and (e)) and the sum of the residual eigenimages (panels (c) and (f)). The expected arrival times, after the group-velocity correction, of the wave packets of the fundamental and the first-higher Rayleigh-wave modes are shown with green and blue dashed lines respectively. The separation appears to be poor especially at the near offsets, where the interference due to the higher mode, contributing to the first eigenimage, seems to deconstruct the signal of the fundamental mode.
Therefore, with the aim of improving our separation technique, we focus on the offset-time portion of the input dataset that contains the fundamental Rayleigh-wave mode only and we attempt its extrapolation in the near-offset portion of the seismogram using the repetitivity characteristic of the first right singular vector of QSVD. To this end, we assume that the experiment is repeated, recording the seismic data also for the offset range 255-500 m where the two modes are better separated, and we select a portion of data containing mainly the fundamental mode. 

The selected portion of data is shown in Fig. \ref{fig:T-16}(a) and (d), for the offset range 5-250 m. As described in Section \ref{se:Q_extraction}, it is possible to extrapolate the missing offsets by extending the first right singular vector $\mathbf{\mathring{v}}_1$ of the QSVD. This allows to estimate the fundamental mode in the near-offset portion, where the two modes interfere. We use this trick to perform the extraction for the frequencies higher than 6 Hz in which both modes propagate (as suggested by the dispersion curves in Fig. \ref{fig:T-8}(a) and (d)). For the low frequencies 4-6 Hz, where only the fundamental mode propagates, we use the whole dataset as input. Fig. \ref{fig:T-16}(b) and (e) show the results of the extraction method on both components using this trick. The residuals, that should contain the higher Rayleigh-wave mode, are shown in Fig. \ref{fig:T-16}(c) and (f). Note that the method satisfactorily reconstructs the fundamental mode in the near-offset portion apart from some noise. This result proves that the right quaternion singular vector $\mathbf{\mathring{v}}_1$ can be used to successfully infer a mode of the Rayleigh wave for missing traces in the seismogram, since it contains information on the offset propagation which is repetitive in the case of laterally-homogeneous vertically-layered media.

\begin{figure}   
  \centering
  \includegraphics[width=1\textwidth]{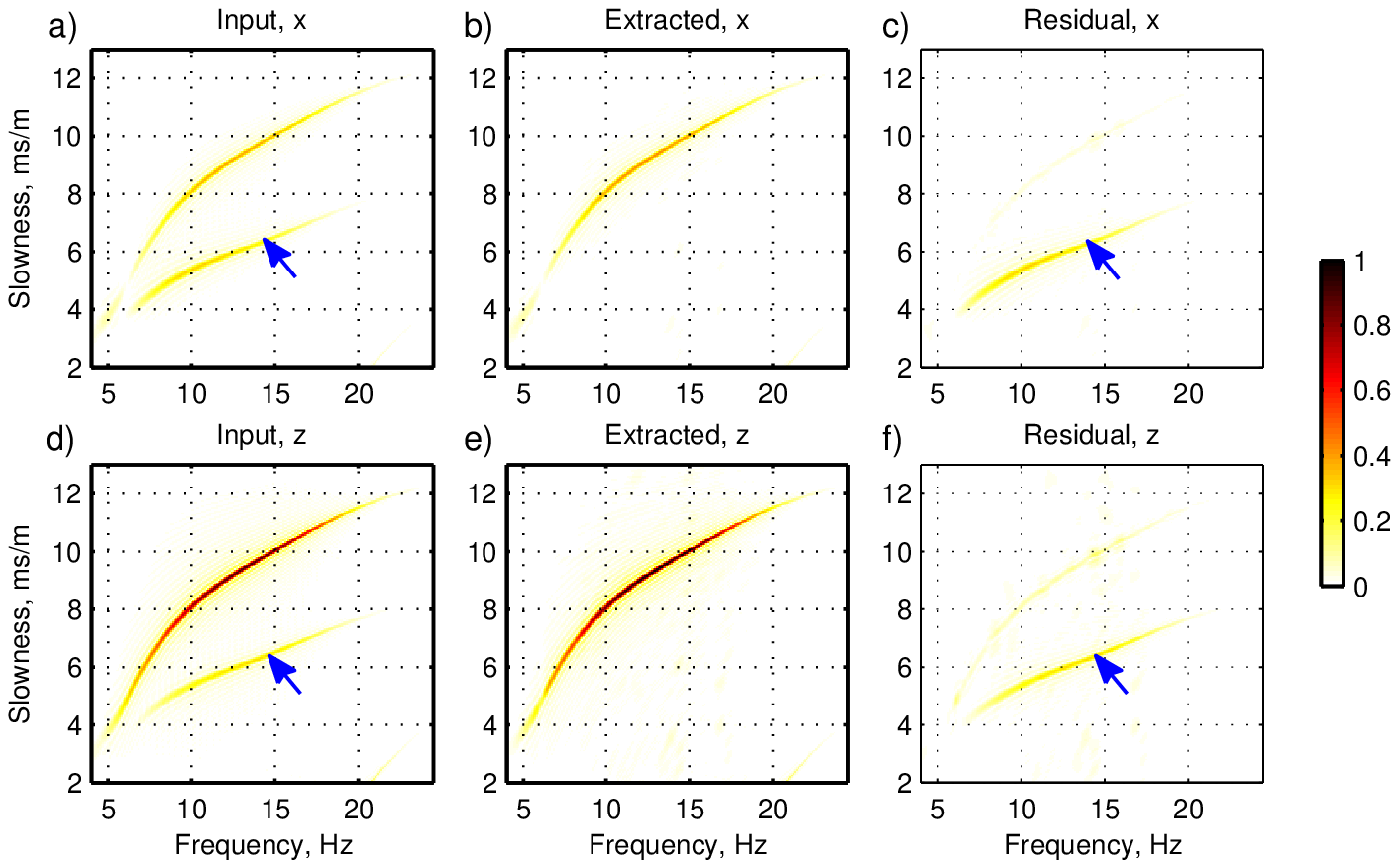} \\
  \caption{(a) f-p transform of the input data, $x$ component; (b) f-p transform of the extracted data, in which the near offsets have been extrapolated using QSVD, $x$ component; (c) f-p transform of the residual data, $x$ component; (d) f-p transform of the input data, $z$ component; (e) f-p transform of the extracted data, in which the near offsets have been extrapolated using QSVD, $z$ component; (f) f-p transform of the residual data, $z$ component. The extracted panel contains almost entirely the fundamental mode with no residual of the higher mode visible.} \label{fig:T-18}
\end{figure}

The results are also demonstrated in the f-p domain in Fig. \ref{fig:T-18}, which compares the f-p transforms of the input ((a) and (d)), of the extracted dataset ((b) and (e)) and of the residual seismograms ((c) and (f)). It can be noted that although some energy of the fundamental mode is visible on the residual panels (c) and (f), virtually no residual of higher Rayleigh wave mode is visible in the extracted panel. These results confirm that identifying a portion of the seismogram in which the mode can be isolated prior to the extraction procedure allows us to satisfactorily reconstruct the missing traces of the mode of interest even where it interferes with a second mode.

\section{Discussion and Conclusions}

We have represented multi-component seismic data by means of quaternions exploiting the properties of Quaternion SVD (QSVD) to describe the inherent vectorial nature of the Rayleigh-wave modes. In particular we have exploited the ability of QSVD to discern rotating aligned signals and take into account simultaneously the dispersion and polarisation effects. Based on these properties, we proposed a procedure that applies appropriate velocity and polarisation corrections to narrow-frequency bands of the seismic data such that the Rayleigh-wave mode of interest behaves like a rotating signal. Written in quaternion notation, this signal is a rank-1 matrix that can be separated from other signals by truncating the QSVD to its first eigenimage. 

We have shown on synthetic data that our procedure successfully recovers the highly-oscillating waveform of the Rayleigh-wave mode. In the application examples included in this work, our procedure has been used to separate this mode from either body waves or a higher mode of the Rayleigh wave. Moreover, we have observed that the first right singular vector after QSVD contains information about the offset dependence of the quasi-monochromatic Rayleigh-wave mode and we have used this property to extrapolate the fundamental mode for missing near-offset traces where the interference of the other mode was affecting the results of our procedure. These findings could in theory be applied to interpolate or extrapolate a mode of the Rayleigh wave for missing traces or for a portion of the dataset, although further studies are required to confirm the reliability and robustness of this procedure. 

For the application results, we have assumed the knowledge of the theoretical values of group velocity and ellipticity characteristics of Rayleigh waves, although methods to estimate these properties are available and seem robust. This has been done because the scope of the work was to demonstrate the correctness of the theoretical model used for representing the single Rayleigh-wave mode. However a number of questions related to the estimation of these parameters remain open. For example it would be interesting to assess how the poor resolution of the group-slowness dispersion curve, especially for some frequencies, might affect the alignment step and therefore the quality of the extraction of the quaternion brick. The ellipticity is the other parameter that strongly affects the circularisation stage and the assumption of rotating signal valid for the Rayleigh-wave mode. 
In this paper we have not discussed in details a method to estimate this parameter although we have compared the theoretical value with an estimate computed as the ratio of the polarisation axes for each frequency of the data. More investigation has to be done in order to understand how to improve this estimation in the presence of multiple modes. Besides, our procedure only works when it recognises the elliptical particle motion of the Rayleigh wave, failing when the narrow-band Rayleigh-wave mode is linearly or quasi-linearly polarised. How to handle these frequencies and how much they affect the quality of the extraction procedure should be further investigated. 

Moreover, due to the complexity of the particle motion, in this paper we focused on the simpler problem of a 2D particle displacement. A 2D particle motion only applies to the special case of sensors correctly aligned along the in-line direction and in the absence of lateral variations in the earth structure. For the special case in which the wave is polarised in a plane defined by the vertical and in-line directions, the formulation can be simplified using complex numbers. However, we show in the paper that the quaternion formulation has the advantage of inherently being able to handle rotations of the plane of motion.

Because we did not have access to real data showing clear elliptical particle motion, we have not tested the method on field datasets. Other practical issues such as the number of receivers, maximum offset, receiver spacing and the effect of spatial aliasing on the estimation of group velocity and more interestingly on the oscillation of the right singular vector are yet to be addressed. We think that these questions, in addition to the effect of estimating the group velocity and the polarisation curves, and the application to real data deserve a separate study.

\section{Acknowledgements}

We thank Alfredo Mazzotti for his suggestions and continuous support. The synthetic data have been generated using the modelling software OASES and the MATLAB$^\circledR$ toolbox \emph{mat\_disperse}, kindly made available by Rix and Lai. We would also like to thank two anonymous reviewers for their interesting remarks.

\section*{References}

\bibliography{mybib_2803}

\appendix

\section{Rotations in Space Using Quaternions} \label{app:quaternion_rotation}

Let $\mathbf{v}$ be a vector in $\mathbb{R}^3$ and $\mathring{v}$ its quaternion representation. We rotate $\mathbf{v}$ by an angle $\phi$ about an axis $\boldsymbol{\omega}$, and we denote with $\mathbf{v}'$ the rotated vector and with $\mathring{v}'$ its quaternion representation. In $\mathbb{H}$, rotations can be written as a combination of a left and a right multiplication by a unit quaternion $\mathring{r}_{\phi/2}$ and its complex conjugate, respectively \cite{LeePhDStanford}:
\begin{equation} \label{eq:appA1}
\mathring{v}' = \mathring{r}_{\phi/2}\mathring{v}\mathring{r}^*_{\phi/2} .
\end{equation} 
where
\begin{equation} \label{eq:appA2}
\mathring{r}_{\phi/2} = \cos(\phi/2) + \sin(\phi/2)\mathring{\omega} ,
\end{equation}
and $\mathring{\omega}$ is the quaternion representation of the rotation axis $\boldsymbol{\omega}$.
We want to demonstrate that, with the assumption made in this paper, that is the motion is constrained into a plane and the axis of rotation is perpendicular to this plane, it also holds that: 
\begin{align} \label{eq:qrot2}
\mathring{v'} &= \mathring{r}_{\phi} \mathring{v} \nonumber \\
\mathring{v'} &= \mathring{v} \mathring{r}_{\phi}^* .
\end{align}
Substituting \eqref{eq:appA2} into \eqref{eq:appA1}, we obtain 
\begin{align} \label{eq:proof_rot1}
\mathring{v}' &= [\cos(\phi/2) + \sin(\phi/2)\mathring{\omega}] \, \mathring{v} \, [\cos(\phi/2) - \sin(\phi/2)\mathring{\omega}] \nonumber \\
&= \cos^2(\phi/2) \mathring{v} + \sin(\phi/2) \cos(\phi/2) \mathring{w} \mathring{v} - \sin(\phi/2) \cos(\phi/2) \mathring{v} \mathring{w} - \sin^2(\phi/2) \mathring{w} \mathring{v} \mathring{w}.
\end{align}
Then we use simple trigonometric rules and the properties $\mathring{v}\mathring{\omega}=-\mathring{\omega}\mathring{v}$ and $\mathring{\omega}\mathring{\omega}=-1$, that hold because $\mathring{v}$ and $\mathring{\omega}$ are perpendicular and pure and $\mathring{\omega}$ is also unitary, to rewrite \eqref{eq:appA1} as
\begin{equation} \label{eq:proof_rot2}
\mathring{v}' = [\cos(\phi) + \sin(\phi)\mathring{\omega}] \, \mathring{v} = \mathring{r}_\phi \mathring{v}.
\end{equation}
A similar proof can be given to show that 
\begin{equation} \label{eq:proof_rot3}
\mathring{v}' = \mathring{v} \mathring{r}^*_\phi.
\end{equation}

\section{Phase Shift as a Rotation} \label{app:phase_shift_as_rotation}

In Sections \ref{se:polarization} and \ref{se:polarization_rot}, we introduced $\underline{u}_{x,m}$ and $\underline{u}_{z,m}$ to represent the horizontal and vertical displacements of a dispersive signal in the narrow-band $[\omega_{0} - \Delta \omega, \omega_0 + \Delta \omega]$ having quasi-circular polarisation and infinite group velocity (see Eq. \eqref{eq:u_xm}). The subscript $m=1,...,M$ denotes the receiver positions. It holds that (see Eq. \eqref{eq:displacement_single_frequency}, Eq. \eqref{eq:xcondition}, Eq. \eqref{eq:constant_phase_shift}, and Eq. \eqref{eq:circularized_displacement}) 
% EQ 12 - EQ 14 -15
% EQ B1
\begin{equation} \label{eq:ipotesi}
\begin{pmatrix} 
\underline{u}_{z,m}
\\ 
\underline{u}_{x,m} \end{pmatrix} =
\begin{pmatrix} 
R_z(t') \sin ( \omega_0 t' - \phi(m) ) 
\\ 
c R_x(t') \cos ( \omega_0 t' - \phi(m) )
\end{pmatrix} .
\end{equation}
Defining
% EQ B2
\begin{equation}
\underline{u}_{m} := 
\begin{pmatrix} 
\underline{u}_{z,m}
\\ 
\underline{u}_{x,m} \end{pmatrix},
\end{equation}
we want to demonstrate that $\underline{u}_{m}$ is a solution of the recurrence relation of Eq. \eqref{eq:rotation_between_columns} that we rewrite here again for readability in a more compact form
% EQ B3
\begin{equation} 
a_{m+1} 
=
P(\Delta \phi)
a_m, \qquad m=1, ... , M-1 ,
\end{equation}
where $P(\Delta \phi)$ is the 2D rotation matrix that represents a rotation about the $y$ axis through an angle $\Delta \phi$
% EQ B4
\begin{equation} 
P(\Delta \phi)=
\begin{pmatrix} \cos (\Delta \phi) & - \sin (\Delta \phi) \\ \sin (\Delta \phi) & \cos (\Delta \phi) \end{pmatrix} \end{equation}
and $a_m$ is a two-component time series $a_m=(a_{z,m},a_{x,m})^T$. One can readily verify that
% EQ B5
\begin{equation} \label{eq:solution}
\tilde{a}_m = 
\begin{pmatrix} 
\sin (\omega_0 t' - m \Delta \phi) \\
\cos (\omega_0 t' - m \Delta \phi) 
\end{pmatrix}
\end{equation}
solves the recurrence relation. The solution $\tilde{a}_m$ is a two-component harmonic signal with circular polarisation and constant phase shift $\Delta \phi$ between adjacent receivers. Note that the solution in Eq. \eqref{eq:solution} is equivalent to the displacement vector of Eq. \eqref{eq:ipotesi} apart from the terms $R_x(t')$ and $R_z(t')$. Hence, if $R_x(t')$ and $R_z(t')$ vary slowly compared to the dominant period $2 \pi / \omega_0$, we can approximate them to be constant $R_x(t') \approx R_x$ and $R_z(t') \approx R_z$. Finally, according to one of the conditions enumerated in Section \ref{se:polarization}, $c R_x \approx R_z$, that is, the horizontal and vertical components have approximately the same amplitude after the polarisation correction. Thus $\underline{u}_{m} \propto a_m , \, \forall m, \, m=1, ..., M,$ solves the recurrence relation.

\end{document}